\documentclass[preprint,nofootinbib]{revtex4-1}
\usepackage{subfigure}   % required only when side-by-side / subfigures are used
\usepackage{graphics, epsfig ,color} % figures en eps
\usepackage{amsmath,amssymb} % environnement American Mathematical Society
\usepackage{enumerate}
\usepackage[dvipsnames]{xcolor}
\usepackage{amsmath,amsfonts,amssymb}
\usepackage{color}
\usepackage{url}
\usepackage{bm}
\usepackage{bbm}
\usepackage{float} 
\usepackage{dsfont} 
\usepackage{mathtools}

\makeindex
\newcommand{\nko}{\tau_k(\omega,n)}

\newcommand{\setZ}{\mathbbm{Z}}
\newcommand{\setR}{\mathbbm{R}}

\newcommand{\Pnc}[2]{\mathds{P}_n\bra{#1 \, \left| \, #2 \right.}}

\newcommand\seq[3]{{#1}_{#2}^{#3}}
\newcommand\sif[1]{\seq{\omega}{-\infty}{#1}}
\newcommand\bloc[2]{{\omega}_{#1}^{#2}}

\newcommand{\ent}[1]{\left[\, #1 \, \right]}

\newcommand{\conv}[1]{\stackrel{#1}{\ast}}
\newcommand{\pare}[1]{\left(\, #1 \, \right)}

\newcommand{\bra}[1]{\left[\, #1 \, \right]}

\newcommand{\Set}[1]{\left\{\, #1 \, \right\}}

\newcommand{\vect}[1]{\pare{\begin{array}{ccc}#1\end{array}}}

\newcommand{\vcF}{\vec{\cF}}
\newcommand{\vcP}{\vec{\cP}}

\newcommand{\vcX}{\vec{\cX}}

\newcommand{\Cell}[2]{{#1}_{#2}} %Cell type, cell index
\newcommand{\V}[2]{V_{\Cell{#1}{#2}}} %Voltage Cell type, cell index
\newcommand{\K}[2]{\cK_{\Cell{#1}{#2}}}  %Kernel Cell type, cell index
\newcommand{\A}[2]{A_{\Cell{#1}{#2}}}
\newcommand{\E}[2]{\cE^{#1}_{#2}}
\newcommand{\F}[2]{F_{\Cell{#1}{#2}}}

\newcommand{\R}[2]{R_{\Cell{#1}{#2}}}
\newcommand{\W}[4]{W^{\Cell{#1}{#2}}_{\Cell{#3}{#4}}}

\newcommand{\Pst}[2]{P_{\Cell{#1}{#2}}}

\newcommand{\cB}{{\mathcal B}}

\newcommand{\cE}{{\mathcal E}}
\newcommand{\cG}{{\mathcal G}}
\newcommand{\cH}{{\mathcal H}}

\newcommand{\cK}{{\mathcal K}}
\newcommand{\cL}{{\mathcal L}}

\newcommand{\cN}{{\mathcal N}}
\newcommand{\cP}{{\mathcal P}}
\newcommand{\cS}{{\mathcal S}}
\newcommand{\cF}{{\mathcal F}}
\newcommand{\cX}{{\mathcal X}}

\begin{document}
%\date{\today}
\title{The Retina as a Dynamical System}
\author{Bruno Cessac}
%\footnote{Author footnote.}}
\affiliation{Université C\^ote d'Azur, Inria, France\\
Biovision team and Neuromod Institute\\
2004 route des Lucioles\\
06901 Sophia Antipolis\\
France.\\
bruno.cessac@inria.fr
%\footnote{Affiliation footnote.}
}

\begin{abstract}
Considering the retina as a high dimensional, non autonomous, dynamical system, layered and structured, with non stationary and spatially inhomogeneous entries (visual scenes), we present several examples where dynamical systems-, bifurcations-, and ergodic-theory provide useful insights on retinal behaviour and dynamics. 
\end{abstract}

\maketitle

%\index[aindx]{Author, F.} % or \aindx{Author, F.}
%\index[aindx]{Author, S.} % or \aindx{Author, S.}

%\markboth{Even Page Header}{Odd Page Header} % Customized running heads

%\body

%\tableofcontents

\section{Introduction}\label{Sec:Intro}

The retina is the entrance to the visual system \cite{besharse-bok:11}\footnote{See e.g. \protect\url{https://webvision.med.utah.edu/}}. The development of new technologies and experimental methodologies (MEA, 2-photon, genetic engineering, pharmacology) and the resulting experiments have made it possible to show that the retina is not a mere camera, transforming the flow of photons coming from a visual scene in sequences of action potentials interpretable by the visual cortex. It appears, in contrast, that the specific and hierarchical structure of the retina allows it to pre-process visual information, at different scales, in order to reduce redundancy and increase the speed, efficiency and reliability of visual responses \cite{gollisch-meister:10}. This is particularly salient in the processing of motion, which is the essence of what our visual system receives as input: there are no static objects since our body, our head, our eyes are constantly on the move. In particular, it has been shown experimentally that the retina can detect a moving object on a background having a distinct motion, anticipate the trajectory of a moving object \cite{gollisch-meister:10} or, even, respond to changes in its trajectory \cite{chen-marre-etal:13}. Such processing are obviously present in the visual cortex, but retinal pre-processing could make visual processing more efficient. For example, by compensating for the delay - mainly due to photo transduction - between the capture of the photons emitted by an object by the photo-receptors and the response of the brain \cite{berry-brivanlou-etal:99}. 

The retina is thus a fascinating object and its study opens the door to better understand how our brain processes information (here, visual information). In this chapter, we adopt a specific, modeller, point of view: The retina is a dynamical system.  This might look, on one hand, quite restrictive, and, on the other hand, not really original. Restrictive because, focusing on a dynamical system perspective, we are somewhat constrained by the structure of equations and the limited tools to analyse them; we are therefore bound to specific problems and questions which necessarily provide a limited view of the retina's behaviour. Not very original because, for sure, the retina is made of neurons, synapses, neurotransmitters, ionic channels, ruled by the laws of physics, written in terms of differential equations. So, yes, retina dynamics can be studied in the realm of dynamical systems theory, but the same is done since long with the brain and cortical neuronal networks models \cite{ermentrout-terman:10}. 
What's new with the retina ?

So, the scope of this chapter is to convince the reader that the retina is a specific organ, with specific neurons and circuits, rather different from their cortical counterparts, leading to specific questions  - requiring specific techniques  - that can be addressed in the realm of dynamical systems theory.  More precisely, the retina is a high dimensional, \textit{non autonomous} dynamical system, layered and structured, with non stationary and spatially inhomogeneous entries (visual scenes). It is therefore necessary to go, as much as we can, beyond the comfortable assumption of stationarity. 

Specific local retinal circuits made of specific neurons, connected with specific non linear synapses, efficiently process the local visual information. Most of the involved neurons - except ganglion cells, the output of the retina - are \textit{not spiking}, in strong contrast with cortical neurons. These local circuits are connected together, spanning the whole retina in a regular tiling, providing parallel channels of information processing \cite{hubel-wiesel:60,barlow:61,olshausen-field:98} that can hardly (from a dynamical system perspective) be considered as independent. Thus, we have to consider, still from a dynamical system perspective, which type of spatio-temporal correlations can be induced by the conjunction of intrinsic neurons dynamics, their interactions by synapses, and their stimulation by a non-stationary drive. This question is particularly salient when the stimulation comes from an object moving along a deterministic trajectory entangling space and time. 

In addition, the retina behaviour can be affected by specific changes in its structure or physiology.  During the early stage of its development, the retina of vertebrates undergo retinal waves, disappearing a few days after birth, which are essential to structure the visual system and make it efficient.  Later in its life, the retina can be affected by specific pathologies, impairing vision. Finally, pharmacology can affect the retinal response to visual stimulation in a controlled way.
These changes and their impact on retina's dynamics can be addressed, too, from the point of view of dynamical systems and bifurcations theory.\\

This chapter presents three examples along these lines. After a brief introduction on retina's structure and dynamics (section \ref{Sec:Retina}), we present, in section \ref{Sec:Waves}, a dynamical system study of retinal waves, published in \cite{matzakou-karvouniari-gil-etal:19,matzakos-karvouniari-cessac-etal:19}. This is an interesting example of multi spatio-temporal scales dynamics where the conjunction of non linearities and closeness to a bifurcation point, combined with a non linear coupling on a regular lattice structure, can produce a highly disordered spatio-temporal dynamics, where the impact of variations of physiological parameters (development, pharmacology) and its consequence on waves dynamics can be addressed. 
In section \ref{Sec:Response}, we consider the effect of lateral amacrine cells connectivity on the response to a moving object, and especially, the anticipation of its trajectory \cite{souihel-cessac:20}. Finally, in section \ref{Sec:Correlations}, we provide a conceptual and operational framework to study the non stationary spike correlations induced by a time dependent stimulus in a neuronal network. Although the related work was initially not specific to the retina \cite{cessac:08,cessac:11,cessac:11b,cofre-cessac:13,cofre-cessac:14} it has resulted in algorithmic tools and a software used to analyse retinal data \cite{cessac-palacios:12,vasquez-marre-etal:12,cessac-kornprobst-etal:17} that we shall not discuss here though, focusing on the mathematical and conceptual consequences of these studies.
The last section tries and generalises the point of view considering  the retina as a dynamical system to more general perspectives.

This chapter mostly present results, published elsewhere, done in collaboration with Ignacio Ampuero, Rodrigo Cofr\'e, Lionel Gil, Dora Karvouniari, Olivier Marre, Serge Picaud, Selma Souihel and I would like to warmly acknowledge them.

\section{Brief overview of the retina} \label{Sec:Retina}

%For more details on the retina structure and function see e.g. \cite{besharse-bok:11} and \url{https://webvision.med.utah.edu/}.

\subsection{Structure}\label{Sec:StructRetina}

The organization of the retina can be found e.g. 
%is shown in the figure \ref{Fig:Retina}.
in \cite{besharse-bok:11} or \url{https://webvision.med.utah.edu/book/part-i-foundations/simple-anatomy-of-the-retina/})). 
It is structured in neuronal and synaptic layers with five neuronal types: Photo-receptors, rod and cones (P), horizontal cells (HCells), bipolar cells (BCells), amacrine cells (ACells), ganglion cells (GCells), to which are added glial cells (Mueller's cells). A feedforward connectivity can be distinguished, the P-B-G path which leads from the photo transduction to the spike trains emitted by the GCells towards the cortex. There is also a lateral connectivity due to Hcells, at the origin of the Center-Surround structure of the receptive fields, and the Acells whose role is still poorly understood and which are one of the main objects of this chapter.

% \begin{figure}
%\centering
%\includegraphics[width=\textwidth,height=0.3\textheight]{RetinaStructureFree.png}
%\caption{\textbf{Structure of the retina. From \protect\url{https://upload.wikimedia.org/wikipedia/commons/thumb/8/88/Retina.svg/2000px-Retina.svg.png?uselang=fr}.
%}
%\label{Fig:Retina}}
%\end{figure}

We can roughly subdivide the retina into two blocks. The first, ending at the OPL (Outer Plexiform Layer), comprises the P, HCells, BCells and the related synapses. The "input" of this block is the flow of photons emitted by the outside world and picked up by the photo-receptors. The "output" is the set of membrane potentials of Bcells, temporally modulated by visual input. More precisely, the voltage of each BCells integrates, spatially and temporally, the local visual information of the photo-receptors which are connected to it, with a lateral modulation due to the HCells. Each BCells is thus sensitive to specific local characteristics of the visual scene, defining its Receptive Field (RF). Mathematically, the receptive field is characterized by the convolution between a spatio-temporal kernel and the stimulus (see section \ref{Sec:ConvKernel}). Nonlinear effects such as gain control must also be taken into account for a realistic characterization of the BCells response. 

The second block is the Inner Plexiform Layer (IPL), comprising the ACells and GCells and the afferent synapses. Its "input" is the output of the OPL, and its output, the trains of action potentials emitted by the GCells, which transmit the flow of information from the retina to the thalamus and the cortex via their axons which constitute the optic nerve. ACells are difficult to study experimentally because hardly accessible from electrophysiology measurements. There are also a large number of cell subtypes in the ACells class
(around 40), of which only a small number have duly identified functions. It is however recognized that they play an essential role in the treatment of motion \cite{baccus-meister:02,nelson-kolb:04,johnston-lagnado:15}.

\subsection{Modelling}\label{Sec:ModellingRetina}

Neurons in the retina have the same biophysics as their cortical counterparts. However, they operate under different regimes. Thus, with the exception of the Gcells, the retinal neurons do not emit action  potentials. %(We are talking here about sodium action potentials, of the Hodgkin-Huxley type). 
Their activity and interactions therefore take place through graded (continuous) membrane potentials as opposed to the sharp peak of an action potential. Furthermore, there is no long-term synaptic plasticity in the retina. Finally, if cortical processing units, such as cortical columns, involve thousands of neurons - making the mathematical description in terms of mean field and neural masses relevant - in the retina the role of these units is played by elementary circuits involving a very small number of neurons, of the order of unity. 

To summarize, the retina is an organ able to perform quite complex tasks in (pre-)processing visual information, with a very low energy consumption, using analogic computing where most neurons do not spike (except GCells). These astonishing capacities are in particular due to specific, local circuits, involving a small number of neurons with specific synapses.

As a consequence, models start from the same biophysical principles as cortical neurons, but the resulting dynamics is different. The mathematical studies can target, as well, different scopes.

\section{The multiscale dynamics of the retina: retinal waves} \label{Sec:Waves}

\subsection{Context}\label{Sec:ContextWaves}

During the early development in the visual system of many vertebrate species \cite{wong-meister-etal:93,feller-butts-etal:97,sernagor-eglen-etal:00, firth-wang-etal:05,warland-huberman-etal:06,sernagor-grzywacz:99,sernagor-hennig:12, ford-feller:12,maccione-hennig-etal:14}, bursts of activity, called \textit{retinal waves}, occur spontaneously and contribute to the visual system organization \cite{wong-meister-etal:93, firth-wang-etal:05, sernagor-hennig:12, ford-feller:12}. They are characterized by localized groups of neurons becoming simultaneously active, initiated at random points and propagating at speeds ranging from $100$ $\mu m/s$ (mouse, \cite{singer-mirotznik-etal:01,maccione-hennig-etal:14}) up to $400$ $\mu m/s$ (chick, \cite{sernagor-eglen-etal:00}), with randomly moving boundaries \cite{ford-feller:12,ford-felix-etal:12}. Wave activity begins in the early development, long before the retina is responsive to light, and it stops a few weeks after birth. It emerges due to several biophysical mechanisms, which changes during development, dividing retinal waves into 3 stages (I, II, III) \cite{sernagor-hennig:12}. Each stage, mostly studied in mammals, is characterized by a certain type of network interaction: gap junctions (electric synapses) for stage I; cholinergic transmission for stage II; and glutamatergic transmission for stage III.

It has been shown that this intermittent spatio-temporal activity  is an inherent property of specific retinal cells and their network interactions \cite{zheng-lee-etal:06}. 
More precisely, the generation of waves at any stage requires three conditions \cite{ gjorgjieva-eglen:11,ford-feller:12}.
(i) A source of depolarization for wave initiation. Given that there is no external input (e.g. from visual stimulation in the early retina), there must be some intrinsic mechanism by which neurons become active; (ii) A network of excitatory interactions for propagation. Once some neurons become spontaneously active, how do they excite neighbouring neurons; (iii) A source of inhibition that limits the spatial extent of waves and dictates the minimum interval between them. \\

In this chapter, we focus on stage II retinal waves where a wide literature exists on the experimental \cite{wong-meister-etal:93,feller-butts-etal:97,sernagor-eglen-etal:00,singer-mirotznik-etal:01, firth-wang-etal:05,warland-huberman-etal:06,sernagor-grzywacz:99,sernagor-hennig:12, ford-feller:12,maccione-hennig-etal:14} 
and modelling side \cite{feller-butts-etal:97,butts-feller-etal:99,godfrey-swindale:07,godfrey-eglen:09,hennig-adams-etal:09,lansdell-ford-etal:14}. The study presented here relies on the works \cite{matzakou-karvouniari-gil-etal:19,karvouniari:18,matzakos-karvouniari-cessac-etal:19} 

During this period, the cells involved are called Starbust Amacrine Cells (SACs) (so-called because of their dendritic tree shape which make them look like the explosion of a star). They interact via the neurotransmitter Acetylcholine (Ach) with nicotinic receptors. Zheng et al \cite{zheng-lee-etal:06} have shown that SACs emit spontaneous intrinsic calcium bursts, which stop after a few hundreds of ms. When 
bursting, a SAC emits Ach, which can in turn excite SACs in its neighbourhood, thereby triggering a wave. Stage II waves are therefore spatio-temporal phenomena resulting from the coupling of local bursters (SACs), spatially coupled by Acetylcholine. In addition, the SAC's bursting generates  a strong after hyperpolarization current (sAHP) which, eventually, drives the SAC to rest, i.e. the bursting phase stops, and induces a long refractory period.
In a network, the propagation of a wave generates therefore  sAHP refractory zones preventing the propagation of a new wave for a period of order a few seconds  \cite{zheng-lee-etal:06}. The sAHP plays thus a prominent role in shaping waves periodicity and boundaries. 

From this brief introduction one may believe that retinal waves show relatively constant characteristics. This is not the case. In contrast they show strong variability. The sizes and durations are  distributed according a power law \cite{hennig-adams-etal:09} and, within stage II, this distribution evolves. One also observes a wide variability in waves inter bursts, period or duration, across different species (e.g. rabbit vs mouse) \cite{ford-feller:12}, but also a variability between individuals within a species. Finally, retinal waves are quite sensitive to pharmacological agents that can increase or decrease their activity. 

In this section, we want to show how these strong variabilities in retinal waves can be explained through a unique mechanism, grounded on dynamical systems analysis and bifurcations theory.

\subsection{The dynamics of Starburst Amacrine Cells}\label{Sec:SACsDynamics}

The details of the model are given in \cite{matzakou-karvouniari-gil-etal:19,karvouniari:18,matzakos-karvouniari-cessac-etal:19}. Here we give the main equations and results.
SACs are considered as points on a lattice, i.e. we do not consider their spatial structure.

\subsubsection{Fast potassium channels}\label{Sec:FastPotassium}

In this model, the fast repetitive firing of SACs is due to the competition between a depolarizing $Ca^{2+}$  current and a fast potassium current. The membrane voltage $V(t)$ of a SAC obeys:

\begin{equation} \label{eq:Voltage1}
C_{m}\frac{d V}{d t}=I_L(V)+I_{C}(V)+I_K(V,N) + I_{ext} + \sigma \xi_t,
\end{equation}
where $C_m$ is the membrane capacitance, $I_L=-g_L(V-V_L)$  is the leak current, with $g_L$, leak conductance and $V_L$, leak reversal potential. The calcium current reads:
\begin{equation} \label{eq:ICa}
I_{C}(V)=-{g}_{C} M_\infty(V)(V-V_{C}),
\end{equation}
where ${g}_{C} M_\infty(V)$ is the voltage dependent conductance of the $Ca^{2+}$ channel, with $M_\infty(V)=\frac{1}{2} [1+\tanh (\frac{V-V_{1}}{V_{2}})]$. 
The \textit{fast} potassium current takes the form:
\begin{equation}\label{eq:IK}
I_K(V,N)=-g_{K}N(V-V_{K}),
\end{equation}
where the evolution of the voltage-gated $K^+$ channel gating variable $N(t)$ is given by:
\begin{equation}\label{eq:N}
\tau_N\frac{d N}{d t}=\Lambda(V)(N_{\infty}(V)-N),
\end{equation}
with $\Lambda(V)=\cosh(\frac{V-V_{3}}{2 V_{4}})$,
and $N_{\infty}(V)=\frac{1}{2} [1+\tanh (\frac{V-V_{3}}{V_{4}})]$, and where $\tau_N$, the characteristic time of the activation variable $N$, is of order $5$ ms.
Although, the form of equations \eqref{eq:IK}, \eqref{eq:N} is based on the Morris-Lecar model \cite{morris-lecar:81}, the variable $N$ is fast, in contrast to the original paper, calibrated in order to capture the frequency of the fast repetitive firing of SACs, (around $20$ Hz \cite{zheng-lee-etal:06}). 

The term $\sigma \, \xi_t$ accounts for noise in the dynamics, where $\xi_t$ is a white noise and $\sigma$ the noise intensity. The role of noise is further discussed below. For the moment, we take $\sigma=0$.\\

\begin{figure}
\centering
\includegraphics[width=10cm,height=8cm]{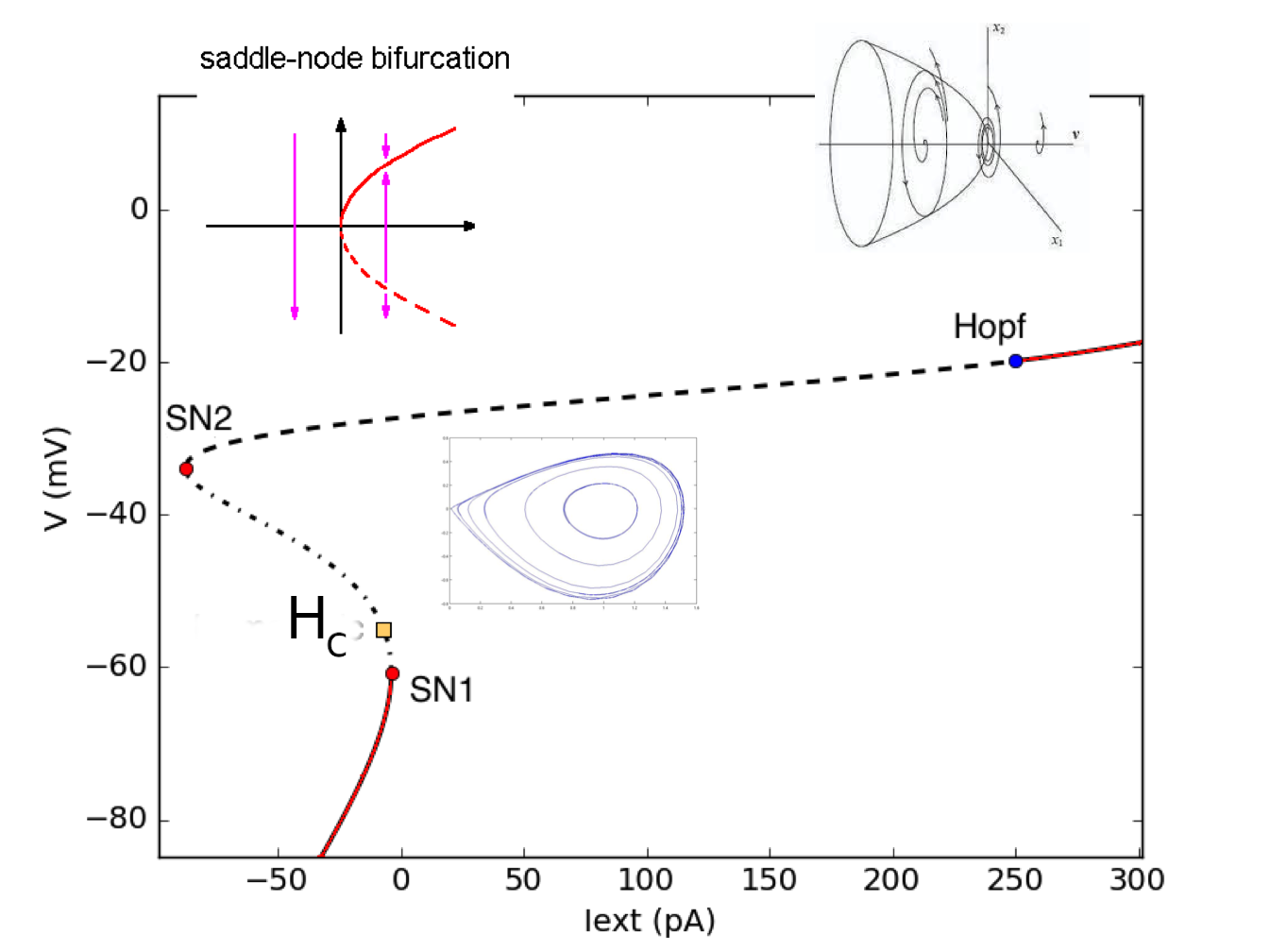}
\caption{\textbf{Bifurcation diagram of \eqref{eq:Voltage1}} in the presence of a constant current ($I_{ext}$). SN stands for "saddle-node". Red full lines correspond to stable fixed points. Small dashed to a saddle-point. Long dashed correspond to a unstable focus. In the inset we have illustrated the typical form of the corresponding bifurcations. See also Fig. \ref{Fig:BifurcationDiagram_GS_GA}. 
\label{Fig:SN_Bifurcation}}
\end{figure}

For the moment too, $I_{ext}$ is considered as a constant, tunable current.
The parameters of the model have, mostly, been tuned from experimental papers (see \cite{matzakou-karvouniari-gil-etal:19} for details). For these parameters we observe the following behaviour when varying the current $I_{ext}$ (Fig. \ref{Fig:SN_Bifurcation}). For low, negative current, the dynamical system admits a unique, stable fixed point (red, full, branch on the left of the diagram). When $I_{ext}$ increases from negative values, there is a saddle-node bifurcation ($SN_2$), at $I_{ext}=I_{SN_2}$, where this fixed point coexists with two unstable fixed points, i.e. an unstable focus (long dashed lines in Fig. \ref{Fig:SN_Bifurcation}) and a saddle (small dashed lines in Fig. \ref{Fig:SN_Bifurcation}). At the homoclinic bifurcation point ($H_c$) ($I_{ext}=I_{H_c}$) a stable limit cycle emerges, with fast oscillations (of order a few ms). The stable fixed point coexists with an unstable focus and this stable limit cycle.  
At the saddle-node bifurcation $SN_1$ ($I_{ext}=I_{SN_1}$) the stable fixed point coalesces with the unstable focus, and only the stable limit cycle  remains.  Note therefore that the homoclinic bifurcation and the saddle-node $SN_1$ are very close together (less than $2$ pA). For simplicity, the next description will do as if they were occurring at the same point although their separation plays an important role in retinal waves distribution of size and durations \cite{matzakos-karvouniari-cessac-etal:19}.
Finally, for a relatively high current ($I_{ext}=I_{Hopf} \sim 250$ pA) the stable limit cycle disappears by Hopf bifurcation giving rise to a stable state. 

Therefore, essentially, going from a negative to a positive current, one switches from a stable rest state with low voltage ($\sim -70$ mV, close to the leak reversal potential $V_L$) to fast oscillations with high average voltage ( $\sim -30$ mV), whereas going from a positive to a negative current, one switches from fast oscillations to stable rest state.

\subsubsection{Slow After-Hypepolarization Current}\label{Sec:sAHP}

In the fast oscillations regime, the average membrane potential has a sufficiently high value to trigger the entrance of calcium via voltage gated calcium channels. This induces a cascade of reactions resulting in the opening of potassium channels and the production of a slow Hyper-polarisation current $I_{sAHP}$. This cascade involves the internal calcium concentration $C$, the fraction of saturated Calmodulin $S$ and the fraction $R$ of bounded terminals in the corresponding potassium channels. The corresponding equations are \cite{matzakou-karvouniari-gil-etal:19}:
\begin{equation}\label{eq:RsAHP}
\left\{
\begin{array}{lll}
\tau_{C}\frac{d C}{d t}&=-\frac{\alpha_{C}}{H_{X}}C+C_{0} + \delta_{C} \, I_C(V);\\
\tau_S \frac{dS}{dt} &= \alpha_S C^4 (1-S) - S;\\
\tau_{R}\frac{d R}{d t}&= \alpha_R \, S(1-R)-R,
\end{array}
\right.
\end{equation}
with the characteristic times $ \tau_{C}\sim 2 \, s$ and $\tau_S,\tau_{R} \sim 40 \, s$. Thus, the calcium dynamics is slow compared to the fast oscillations (time scale of order a few ms) while the variables $R,S$ are slow compared to calcium dynamics. The other parameters are fixed from biophysics \cite{karvouniari:18}.

This cascade of reactions induces a sAHP current of the form 
\begin{equation}\label{eq:IsAHP}
I_{sAHP}(V,R)=-g_{sAHP}R^4 \, (V-V_{K})  \equiv -G_S \, (V-V_{K}) ,
\end{equation}
where $g_{sAHP}$ is the maximum sAHP conductance and $G_S=g_{sAHP}R^4$ is the effective conductance. It involves the probability, $R$, that one terminal of a slow potassium channel is bounded. This probability has the power $4$ because $4$ bound terminals are needed to open a $Ca^{2+}$-gated $K^+$ channel. 

Since the time variation of $G_S$ is extremely slow compared to the fast oscillations of \eqref{eq:Voltage1} one can average over the fast term $V-V_{K}$ in \eqref{eq:IsAHP}, and consider $I_{sAHP}$ as a slow varying parameter, corresponding to $I_{ext}$ in the bifurcation diagram \ref{Fig:SN_Bifurcation}. (See also Fig. \ref{Fig:BifurcationDiagram_GS_GA} for a bifurcation diagram dealing with the slow variables).

So, the picture is the following. In the fast oscillations regime, $V$ is high, thereby triggering the increase (in absolute value) of a slow negative (potassium) current $I_{sAHP}$ eventually leading the system to cross the bifurcation point $SN_1$. Here, oscillations stop. Because, sAHP is slow there is some inertia where $I_{sAHP}$ goes on decreasing, hyperpolarizing the neuron. In this stage, calcium entrance stops, so, eventually, the $I_{sAHP}$ will become less and negative and returns back to its rest value $I_{sAHP}=0$.
The corresponding periodic variation of the variables $C,R,S$ is shown in Fig. \ref{Fig:BifurcationDiagram_GS_GA} bottom, right.

Here, actually, two situations are possible. In the case depicted in Fig. \ref{Fig:SN_Bifurcation}, $I_{SN_1}>0$, so when 
$I_{ext}=I_{sAHP}=0$, fast oscillations restart and we are back to the initial situation. Thus, we have a simple mechanism explaining the period bursting of cells (alternation of fast oscillations and rest state) observed in experiments. We call this case spontaneous bursting (SB).

By a slight change of parameters, it is also possible to make the $SN_1$ and homoclinic bifurcation arising for a strictly positive value of $I_{ext}$. Then,  $I_{ext}=I_{sAHP}=0$ corresponds to the rest state on the stable branch left-wise to $SN_1$. If, however, the $SN_1$ bifurcation is close to $I_{ext}=I_{sAHP}=0$ ($I_{SN_1} \stackrel{<}{\sim} 0$) then it is possible to restart oscillations with a bit of noise (the term $\sigma \xi_t$ in \eqref{eq:Voltage1}). Then, the same process as above occurs, but, now, the sequence of bursts is not periodic any more.  It occurs on random times. We speak of Noise Induced Bursting (NIB).

In the NIB regime the average time between two bursts depends strongly on the closeness of $I_{SN_1}$ to $0$. As shown in \cite{matzakou-karvouniari-gil-etal:19} this sensitivity provides a simple mechanism to explain the variability in inter burst intervals observed across and within species.

\subsubsection{Acetylcholine coupling}\label{Sec:Ach}

In this stage of development SACs are interacting via the neurotransmitter Acetylcholine  (Ach) with nicotinic receptors. We consider from now a network of $N$ SACs, with index $i=1 \dots N$, organised on a regular lattice (e.g. nearest neighbours or a connectivity that mimics the biological reality \cite{matzakos-karvouniari-cessac-etal:19}) in one, or two dimension. 

The concentration $A_j$ (in nM) of Ach  emitted by a pre-synaptic cell $j$, with voltage $V_j$, is given by the differential equation:
\begin{equation} \label{eq:Achprod}
\frac{d A_j}{d t}=-\mu \, A_j \, + \, \beta_A T_A(V_j),
\end{equation} 
where $\mu$ is a degradation coefficient, $\beta_A$ is the maximal production rate of $A$ (in $nM s^{-1}$), and the production term:
\begin{equation}\label{eq:tach}
T_A(V_j)=\frac{1}{1+e^{-\kappa_A (V_j-V_A)}},
\end{equation}
where $V_A$ is the Nernst potential associated to Ach.
The parameters $\mu,\kappa_A,\beta_A,V_A$ have been fitted from experiments\cite{zheng-lee-etal:06,matzakos-karvouniari-cessac-etal:19}. They are uniform on the SACs population (independent of $j$).

The Ach  released in the synaptic cleft from pre-synaptic neuron $j$ to post-synaptic neuron $i$ binds to 2 nicotinic receptors triggering the opening of specific ionic channels. This induces, in the post synaptic neuron $i$ an excitatory current
$ -g_A \, (V_i-V_A) \, \frac{A_j^2}{\gamma_A+A_j^2}$.
Let us denote $\cB_i$ the neighbourhood of $i$, that is, the pre-synaptic neurons connected to $i$ by cholinergic coupling.
The total cholinergic current injected in neuron $i$ is:
\begin{equation}\label{eq:IA}
I_{A_i} = -g_A \, (V_i-V_A) \, \displaystyle{\sum_{j \in {\cal B}_i}}{\, \frac{A_j^2}{\gamma_A+A_j^2}} \equiv -G_A \, (V_i-V_A),
\end{equation}
where $g_A$ is the maximal Ach conductance and $G_A=g_A \, \displaystyle{\sum_{j \in {\cal B}_i}}{\, \frac{A_j^2}{\gamma_A+A_j^2}}$ the effective conductance. The evolution of $G_A$ is slow compared to the fast oscillations ($\sim$ a few seconds). 
 
The Ach current induces thus a non linear coupling between cells.

\subsection{Retinal waves propagation}\label{Sec:WavesPropagation}

We have now the 3 ingredients to trigger, propagate and control waves: fast bursting, mutual excitation and slow hyper polarisation. Let us see how these mechanism collaborate or compete to produce the wide variability in retinal waves dynamics.

\subsubsection{Toward a generic mechanism controlling waves}\label{Sec:GenMech}

In general, thus, a SAC is subject to two competing current: the (extremely slow) hyper-polarizing current sAHP \eqref{eq:IsAHP} and the (slow) depolarizing cholinergic current \eqref{eq:IA}. 
One easy way to summarize the competition between these two slows and competing mechanism is to draw the equivalent of the bifurcation diagram Fig. \ref{Fig:SN_Bifurcation} of \eqref{eq:Voltage1} in a plane $G_S,G_A$. This way, $G_S,G_A$ become two free parameters that we can vary independently. 

\begin{figure} 
\centerline{
\includegraphics[width=16cm, height=11cm]{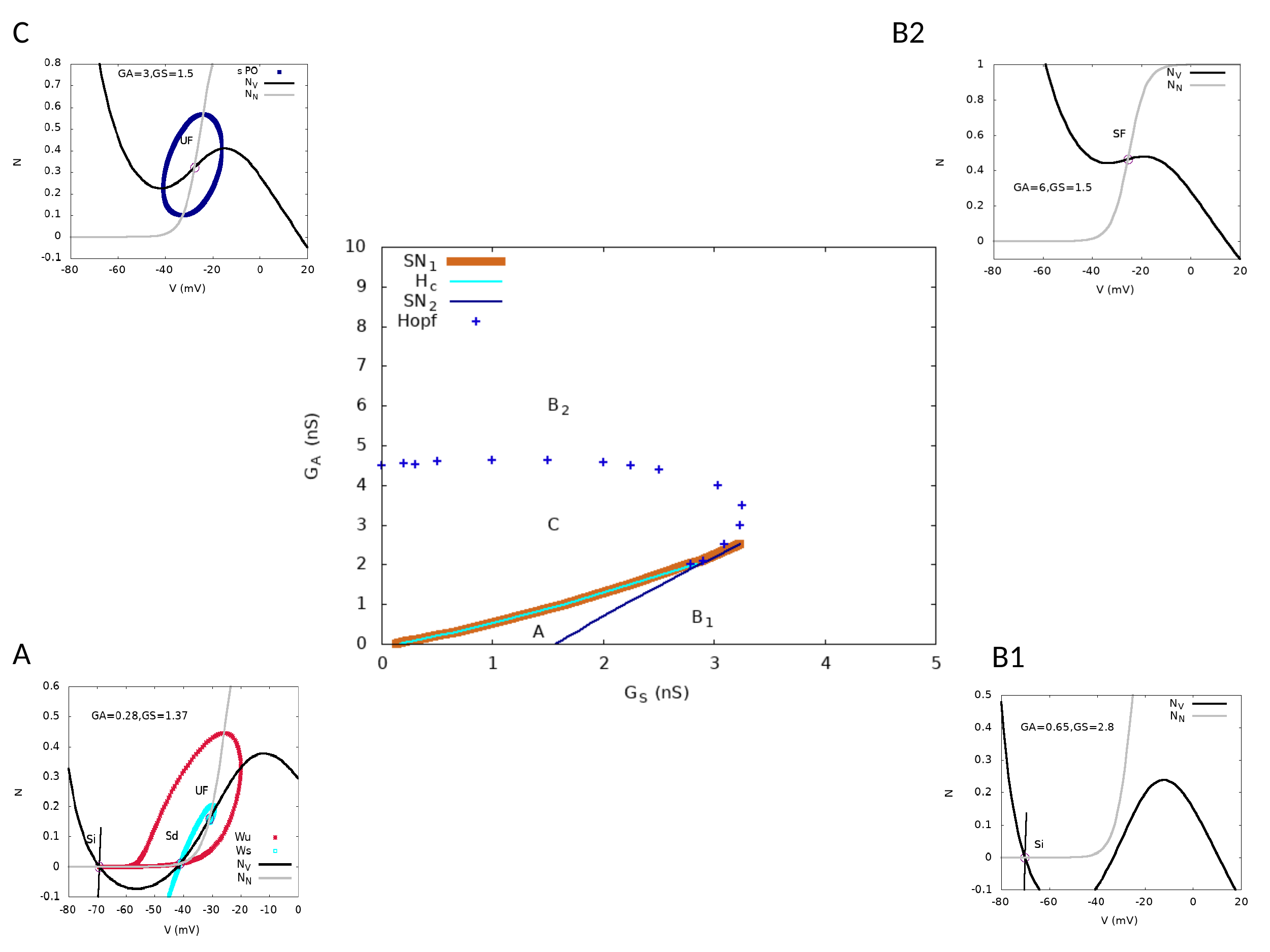}
}
\vspace{0.1cm}
\centerline{
\includegraphics[width=7cm, height=5cm]{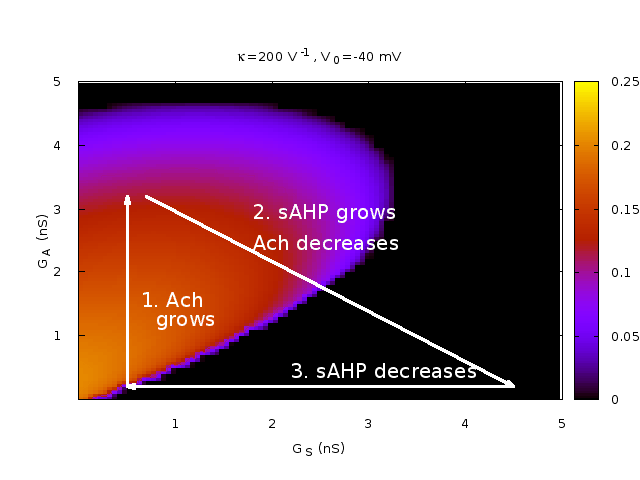}
\hspace{2cm}
\includegraphics[width=7cm, height=5cm]{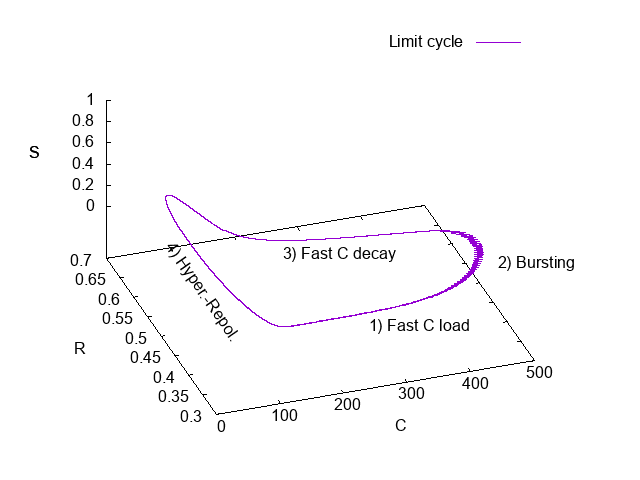}
}
 \caption{\footnotesize\textbf{The synthesized  mechanism of waves dynamics. Top, Center.} Bifurcation diagram of the fast dynamics \eqref{eq:Voltage1} in the plane $G_S$ (effective sAHP conductance), $G_A$ (effective Ach conductance). The panels (A,$B_1$,$B_2$,C) around the bifurcation map are the phase portrait with nullclines ($N_V$, in black, is the $V$ nullcline, $N_N$, in grey, is the $N$ nullcline), stable manifold of hyperbolic fixed points ($W_s$ in blue, panel A), unstable manifold of the saddle-fixed points ($W_u$ in red, panel A), stable periodic orbit (in blue, panel C). "$S_d$" means "Saddle", "SF" means "Stable Focus", "UF" means "Unstable Focus", "SPo" means "Stable Periodic orbit". \textbf{Bottom, Left.} Heat map of the voltage fluctuations in the $G_S,G_A$ map. 
 The colored region corresponds to region C in the bifurcation map. The white arrow sketches the typical evolution of $G_S,G_A$ for a cell embedded in a wave. If the cell is at rest (region $B_1$ or $A$) the increase in Ach current leads it to region C where this cell bursts. It starts to produce sAHP so $G_S$ increases. At some time $G_A$ starts to decrease when neighbouring SACs stop bursting. Although the increase in $G_S$ and the decrease in $G_A$ are independent, we sketch here the compound effect along a straight line in the plane $G_S,G_A$. At some time, the cell leaves region C and returns to region $B_1$. Then, sAHP decreases and the cell returns to its initial state. \textbf{Bottom, Right.} Typical course of the variables $C,R,S$ for a cell. "Hyper-Repol" means "Hyperpolarization-Repolarization". \label{Fig:BifurcationDiagram_GS_GA}}
 \end{figure}

The complete picture is represented in Fig. \ref{Fig:BifurcationDiagram_GS_GA} (see legend for detail). On the top figure we see the bifurcation diagram of \eqref{eq:Voltage1} in the plane $G_S,G_A$. The region of fast oscillations is region $C$. This is visible in the heat map of Fig. \ref{Fig:BifurcationDiagram_GS_GA}, Bottom, Left, where the color represents the amplitude of oscillations (no oscillation in the rest state, corresponding to black regions). In the same figure, we have sketched a typical pathway showing the competition between sAHP and Ach current and the resulting dynamics. Finally, in  Fig. \ref{Fig:BifurcationDiagram_GS_GA}, Bottom, Right we show the typical evolution of the variables $C,R,S$ determining the sAHP conductance, in a cycle where bursting alternates with slow hyperpolarization. There is a remarkable line in the bifurcation diagram: the line where $SN_1$ (red) and $H_c$ (blue) are superimposed. Although they are not identical, they are very close (see also Fig. \ref{Fig:SN_Bifurcation}) and are indistinguishable at this scale of $G_S,G_A$. As already explained when commenting Fig. \ref{Fig:SN_Bifurcation} we will consider that these two bifurcations occurs at the same point. It corresponds to the switch from stable rest state to fast oscillations.

When dealing with a network of SACs, one observes spatio-temporal variations of $G_S,G_A$, organized in regions of excitable cells, or refractory cells. These regions evolve in time. When a SAC starts to burst it recruits, via Ach coupling, excitable cells, which in turns can recruit other cells, generating a wave, propagating until the front meets a refractory zone with high sAHP, corresponding to cells which have recently undergone a wave. Then, what determines if the wave propagates to the next cells is a threshold condition directly related to the $SN_1-H_c$ bifurcation. This results in a high spatio temporal variability where waves propagates in a sAHP landscape, itself slowly evolving in time, and driven by waves dynamics. 

There is some analogy with forest fires \cite{hennig-adams-etal:09,lansdell-ford-etal:14}, although the detailed dynamics of SACs wave is a bit more subtle as we now discuss.  

\subsubsection{Chain of SACs}\label{Sec:ChainSacs}

We start from a one-dimensional chain of $N$ SACs and we consider first the case ... $N=2$.
The first cell, $C_1$, starts to burst, and, if $G_A$ is large enough, induces the second cell, $C_2$ to burst. Then, $C_2$ generates an Ach current which \textit{prolongs} the burst of $C_1$ and increases the peak in Ach production for the two cells. Both cells are then mutually coupled and this mutual coupling reinforces their synchrony. Here, we have assumed that all cells have the same time constant and we only observe 1:1 locking, but, in a population of heterogeneous cells it is possible to exhibit m:n synchronies in patterns corresponding to Arnold tongues \cite{arnold:83,karvouniari-gil-etal:16}. We will not develop on this situation here.
The 2 SACs synchrony lasts until $I_{sAHP}$ is large enough to stop bursting in both cells, but not necessarily at the same time.  

From these considerations it is possible to compute the time $t_B$ when the cell $C_2$ starts to burst if the cell $C_1$ has started to burst at time $0$ \cite{karvouniari:18}. This time depends on model's parameters fixing the value of $I_{SN_1}$ and Ach production. It also depends on the value of the refractory variable $R$ at time $t_B$, illustrating, once again the delicate balance between $I_{sAHP}$ ($G_S$) and $I_A$ ($G_A$).

\begin{figure} 
\centerline{
\includegraphics[width=0.4\textwidth, height=6cm]{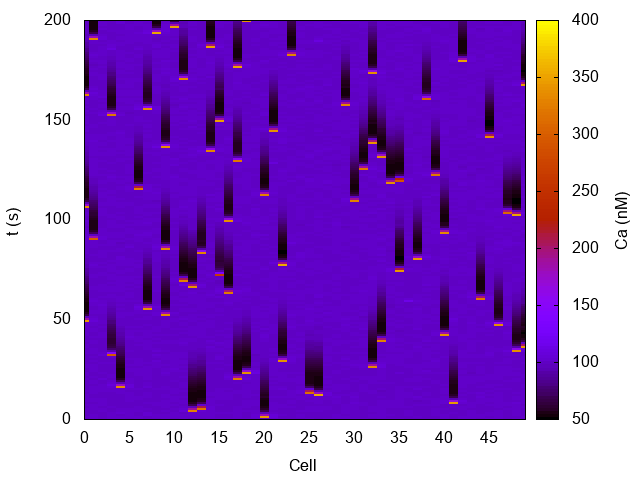}
\hspace{0.1cm}
\includegraphics[width=0.4\textwidth, height=6cm]{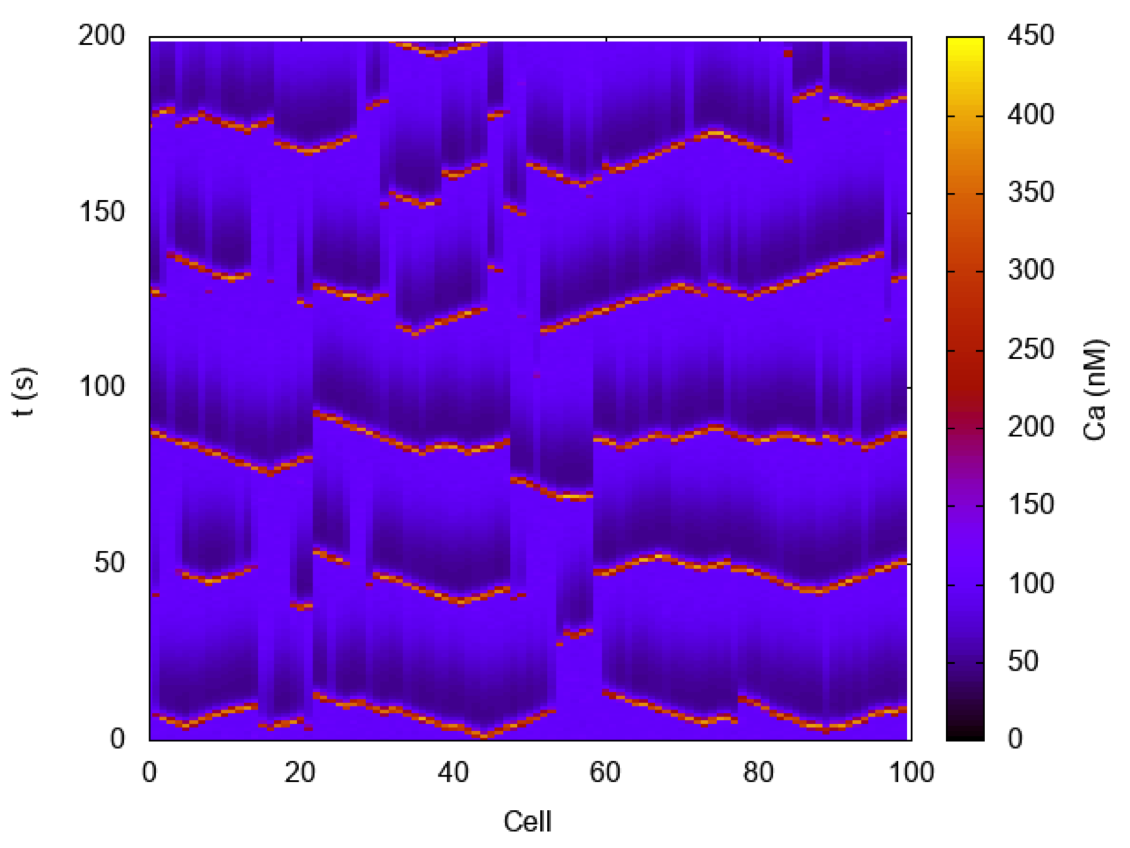}
\hspace{0.1cm}
\includegraphics[width=0.4\textwidth, height=6cm]{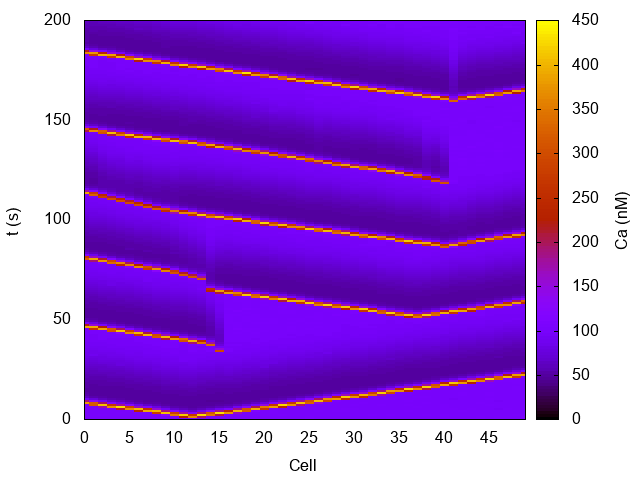}
}
 \caption{Example of propagation of waves, near the $g_A$ critical point $g_{A_c}$, in the presence of noise ($\sigma=3  \, pA \, ms^{-\frac{1}{2}}$) for $50$ cells.   
 Left: $g_A=0.04$ nS; Middle: $g_A=0.07$ nS ; Right: $g_A=0.1$ nS. Red zones are active (bursting) zones; black zones are hyperpolarized (sAHP) zones; purple zones correspond to SACs at rest. In the middle figure, red zones (propagating waves) can be stopped by hyperpolarized regions. 
 \label{Fig:PropagationCriticalNonInteract}. }
 \end{figure}

 It is also possible to show, that, in the NIB regime ($I_{SN_1}>0$) there is a minimal value $g_{A_c}$ for $g_A$, the maximal conductance, below which $C_1$ cannot excite $C_2$. This observation is fundamental to understand the propagation of wave in a one dimensional chain of $N$ SACs. In Fig. \ref{Fig:PropagationCriticalNonInteract} we show the initiation and propagation of waves for increasing $g_A$. On the left, the entrance of a SAC in the bursting regime initiates a propagation which ends very soon. This is because, for this regime of parameters, the Ach current, even at its maximum is not sufficient to cross the SN barrier. Only noise affords it, with a low probability, depending on noise intensity and distance between $g_A$ and $g_{A_c}$. On the opposite, on the right-most figure we see waves essentially propagating through the whole lattice. This is because the sAHP barrier is not sufficient to stop waves propagation. Therefore, there is an intermediate regime (Fig. \ref{Fig:PropagationCriticalNonInteract}, middle) where waves propagation has to cope with the sAHP landscape left by the previous waves, and where there is a wide spread in the distribution of waves (power laws).\\

One can show that there is a characteristic correlation length, corresponding to the average size of active regions and depending on
 the parameters $g_{sAHP}$ and $g_A$ \cite{karvouniari-gil-etal:16} . In addition, near $g_{A_c}$, one observes power laws in the waves distribution (size and duration) and the correlation length increases  \cite{matzakos-karvouniari-cessac-etal:19}. At high $g_A$ waves spread along the full lattice and the correlation length is the lattice size. 
 
\subsubsection{Two dimensional dynamics}\label{Sec:TwoD}

The situation in two dimensions is very similar. The main difference is the topology of wave fronts, which are points in one dimension and becomes complex lines in two dimensions. Therefore, it is quite more difficult to determine the wave shape, especially when two waves interact, rendering difficult a statistical analysis of waves statistics. Example of two dimensional simulations (made by Lionel Gil) can be found at \url{https://www.youtube.com/watch?v=shMR3NMCBDE}. 

A big difficulty in simulating the two dimensional case is the large time scale separation between dynamics. To simulate the complete dynamics one has to use an integration scheme with a very small step, lower than one millisecond, while the time scale of duration of sAHP regions last of order minutes. Therefore, it requires a very long time to simulate a large number of waves to get their statistics.

At this stage, it would certainly be more useful to derive, from this model, another model averaging over the fast dynamics to produce transport equations correctly accounting for waves initiation, propagation and stop. Some attempts in this direction has been made \cite{feller-butts-etal:97,gjorgjieva-eglen:11,godfrey-swindale:07,hennig-adams-etal:09,lansdell-ford-etal:14} but none of them really take into account the closeness to bifurcation, at the core of the wide variability in waves characteristics (see section \ref{Sec:Variability}), where one would like to have, in addition, a scaling theory to describe waves statistics when approach the critical point. A potential candidate is a Kardhar-Parisi-Zhang-like  equation for retinal waves taking into account the spatio-temporal variations of sAHP (in preparation and \cite{cessac:18}).

\subsection{Consequences}\label{Sec:ConsequencesWaves}

We would like now to shortly propose some potential consequences of this analysis.

\subsubsection{Explaining variability with a unique mechanism}\label{Sec:Variability}

As announced in the introduction of this section the wide variability in waves distribution can be explained by the closeness to a saddle-node bifurcation point. 
The spatio-temporal variability in waves distribution for a given set of parameters have been explained throughout this section. Now, during development, and even in the stage II phase, the maximal Ach conductance $g_A$ is known to evolve -  \cite{zheng-lee-etal:04}, Zheng et al have shown that the intensity of the Ach coupling is \textit{monotonously decreasing} with time -  thereby inducing strong variations in the waves evolution.
We have also seen that the closeness to SN in the NIB regime leads to strong variations in the interburst interval. This could explain the variability observed across and within species \cite{matzakou-karvouniari-gil-etal:19}. Finally, 
one interest of our model is to take into account physiological parameters that can be modified upon pharmacology, inducing potential bifurcations. Some of them are studied in \cite{matzakou-karvouniari-gil-etal:19}.

More generally, the variations in  physiological parameters can be induced by: (i) development; (2) pharmacology; (3) pathologies. Indeed, during retinal diseases such as Retinitis Pigmentosa or Age Macular Degeneration, the structure and dynamics of the retina is impacted inducing, in addition to the loss of sight, spatio-temporal phenomena such as large scales oscillations that could be studied with the same type of analysis made here (local bifurcation and extension to network) although the involved cells are not SACs \cite{choi-zhang-etal:14}.  

\subsubsection{Dynamical response to stimuli}\label{Sec:DynRep}

Waves distribution can exhibit power laws (in our model, close to the critical line between A and C in Fig. \ref{Fig:BifurcationDiagram_GS_GA}). This was first observed by M. Hennig et al, in a paper combining experiments and modelling. Investigating the dynamics of stage II retinal waves, they show that the network of SACs is capable of operating at a transition point between purely local and global functional connectedness, corresponding to a percolation phase transition, where waves of activity - often referred to as "avalanches"- are distributed according to power laws (see Fig. 4 of \cite{hennig-adams-etal:09}). They interpret this regime as an indication that early spontaneous activity in the developing retina is regulated according to the following principle; maximize randomness and variability in the resulting activity patterns. This remark is in complete agreement
with the idea of dynamic range maximization \cite{ribeiro-copelli:08} and could be of central importance for our understanding of the visual system. In fact, it suggests that, during its formation, the visual system could be driven by spontaneous events, namely the retinal waves, exhibiting the characteristics of a second order phase transition. However, giving solid evidence that a second order phase transition occur in this model, namely satisfies the relations between critical exponents predicted e.g. by the renormalisation group is a hard task\cite{sethna:95,ma:01,volchenkov-blanchard-etal:02,ribeiro-copelli:08}, even numerically, because of the large range in time scales variations in the dynamics, requiring prohibitive simulations times (especially in two dimensions) to obtain a sufficiently large sample of waves. There might be more hope to develop a model where fast time scale is averaged out \cite{lansdell-ford-etal:14}. To our best knowledge, no renormalisation group analysis has been yet in models of retinal waves.  

It has been suggested \cite{haldeman-beggs:05,shew-plenz:13} that such the scale-free organisation observed in many neural systems \cite{beggs-plenz:03,shew-plenz:13} could foster information storage and transfer, improvement of the computational capabilities \cite{bertschinger-natschlager:04}, information transmission \cite{beggs-plenz:03,bertschinger-natschlager:04,shew-yang-etal:11}, sensitivity  to  stimuli  and  enlargement  of  dynamic  range  \cite{kinouchi-copelli:06,shew-yang-etal:09,gautam-hoang-etal:15,girardi-schappo-bortolotto-etal:16}. In this spirit, retinal waves could certainly play a role in the organization of the visual system to a point with maximal efficacy in response to the wide spatio-temporal range of visual stimuli. 

\section{Response to stimuli}\label{Sec:Response}

Many branches of dynamical system theory, many dynamical system studies, deal with stationary (time translation invariant) situations. This is somewhat similar in the study of neuronal models where entries are mostly static (constant or white noise). Unfortunately, the retina (and thereby the visual cortex) receives constantly non stationary entries. Indeed, \textit{in vivo}, the retina has to cope with moving objects in a moving world. 

These remarks open up the necessity to adapt or extend theoretical methods of dynamics analysis to non stationary situations. To our best knowledge, such methods are at an early stage of development, and the reader shouldn't expect to find anything exceptional on this topic in these pages. Nonetheless, a few interesting results can be obtained, enlightening the potentialities of the retina. In this section, we give an example of non stationary response of the retina, in the context of retinal anticipation. 
We use the fact that bipolar and amacrine cells are not spiking to propose an approximation based on piecewise linear dynamics where non stationarity can be handled. This allows us to predict a potentially new effect of lateral amacrine connectivity, fully studied in \cite{souihel:19,souihel-cessac:20}: the possibility that a stimulus, moving on a deterministic trajectory, triggers a wave of activity, \textit{ahead} of the stimulus,
thereby allowing the retina to anticipate motion, enhancing local non linear gain control mechanisms reported in \cite{berry-brivanlou-etal:99,chen-marre-etal:13}, and shortly introduced below.

\subsection{Convolution kernel}\label{Sec:ConvKernel}

The "first" stage of the retina, ending at the OPL, including photo-receptors, Hcells and Bcells is often represented by a kernel, integrating the processing of these different layers and characterizing the so-called Receptive Field (RF) of a Bcell. In this description, the voltage of a Bcell $i$ is modulated, stimulus-driven, by the term:
\begin{equation}\label{eq:Vdrive}
V_{i_{drive}}(t)= \bra{\K{B}{i} \conv{x,y,t} \cS}(t) =
\int_{x=-\infty}^{+\infty} \,\int_{y=-\infty}^{+\infty} \,
\int_{s=-\infty}^{t} \,  \cK(x-x_i,y-y_i,t-s) \, \cS(x,y,s) dx \, dy \, ds,
\end{equation}
where $\conv{x,y,t}$ means space-time convolution.

On biological grounds, a Receptive Field (RF) is a region of the visual field (the physical space) in which stimulation alters the voltage of the Bcell.  Mathematically, this is, typically, the product\footnote{More generally, spatio-temporal kernels are not separable in the product of a space- and a time-dependent function. The mathematical analysis described here   actually does not require this separability but we choose this form for simplicity.} $\cK(x,y,t)=\cK_S(x,y) \, \cK_T(t)$. Here, 
\begin{equation}\label{eq:KS}
 \cK_S(x,y) = \frac{A_1}{2 \pi \sqrt{\det C_1}}\, e^{-\frac{1}{2} \, \tilde{X_i}.C_1^{-1}.X_i} \,-\,  \frac{A_2}{2 \pi \sqrt{\det C_2}}\, e^{-\frac{1}{2} \, \tilde{X_i}.C_2^{-1}.X_i},
\end{equation}
is the spatial part of the RF. This is a difference of (oriented) Gaussian that mimics the center-surround organization of RF \cite{hubel-wiesel:62},
where $X_i=\vect{x-x_i\\y-y_i}$, $\, \widetilde{} \,$ denotes the transpose, $x_i$ and $y_i$ are the coordinates of the receptive field center which coincide with the coordinates of the cell, $C_1, C_2$ are positive definite matrix whose main principal axis represent the preferred orientation of the RF. The two Gaussians of the RF are thus concentric. 
The temporal part of the RF, $\cK_T$, can have different form, but, it decays sufficiently fast (e.g. exponentially) as time $t \to +\infty$. Typically, this is the difference of decaying exponentials.

In our model, the BCell voltage is the sum of the external drive \eqref{eq:Vdrive} received by the BCell and of a post-synaptic potential $\Pst{B}{i}$ induced by connected ACells:
\begin{equation}\label{eq:VBipTot}
\V{B}{i}(t)=V_{i_{drive}}(t)  + \Pst{B}{i}(t).
\end{equation}
The form of $\Pst{B}{i}$ is given by eq. \eqref{eq:PSP_A_Bip} below. $\Pst{B}{i}(t)=0$ when no ACells are considered.

\subsection{Non linearities}\label{Sec:NonLinearities}

BCells have voltage threshold \cite{berry-brivanlou-etal:99}:  
\begin{equation} \label{eq:NLinRectBip}
\cN_B(\V{B}{i}) =\left\{
	\begin{array}{ll}
		0, \quad &\mbox{if} \quad \V{B}{i} \le \theta_B; \\
		\V{B}{i}-\theta_B, \quad &\mbox{else},
	\end{array}
	\right.
\end{equation}
and gain control, a desensitization when activated by a steady illumination \cite{yu-sing-lee:05}, mediated by a rise in intracellular calcium $Ca^{2+}$, preventing thus prolonged activity \cite{snellman-kaur-etal:08,chen-marre-etal:13}. Gain control is associated to an activity variable $\A{B}{i}$ obeying the differential equation:
\begin{equation}\label{eq:dA}
\frac{d\A{B}{i}}{dt} =  -\frac{\A{B}{i}}{\tau_a} + h_B \, \cN(\V{B}{i}(t)).
\end{equation}

The bipolar output to ACells and GCells is then characterized by a non linear response to its voltage variation,
 given by : 
\begin{equation} \label{eq:ResponseBip}
\R{B}{i}\pare{\V{B}{i}, \A{B}{i}}=
\cN_B\pare{\V{B}{i}} \, \cG_B\pare{\A{B}{i}},
\end{equation}
where :
\begin{equation} \label{eq:gain_control_bip}
\cG_B(\A{B}{i})=\left\{
	\begin{array}{ll}
		0, & \mbox{if} \hspace{0.2cm} \A{B}{i} \le 0; \\
		\frac{1}{1+\A{B}{i}^6}, & \mbox{else}. 
	\end{array}
	\right.
\end{equation}
Gain control plays an important rule in regulating neurons activity, as it   
adjusts the relationship between dramatically changing inputs and the limited range of amplitudes in the neuron's response of neurons \cite{shapley-victor:81}\cite{priebe-ferster:02,jarvis:18}. It has been shown in \cite{berry-brivanlou-etal:99} that gain control in BCells and GCells allows the retina to anticipate the trajectory of a moving object, by shifting its peak of activity before the peak obtained without gain control. Mathematically, this anticipation effect is obtained thanks to eq. \eqref{eq:dA},\eqref{eq:ResponseBip}, \eqref{eq:gain_control_bip}. When a BCell is submitted to a visual stimulation its voltage $\V{B}{i}$ increases\footnote{For simplicity, we do not distinguish here between ON or OFF cells and on the nature of the stimulus.}, thereby increasing the activity $\A{B}{i}$. When $\A{B}{i}$ becomes sufficiently large, $\cG_B(\A{B}{i})$ drops down, hence $\R{B}{i}$ drops down before the voltage $\V{B}{i}$ drops down, thereby advancing the peak of activity.

\subsection{Network}\label{Sec:Network}

We consider a regular lattice of BCells and a regular lattice of ACells, with the same number of cells.
ACells are connected to BCells via a connectivity matrix $\W{A}{}{B}{}$ such that $\W{A}{j}{B}{i} \leq 0$ is the inhibitory\footnote{Most ACells are inhibitory, glycinergic or GABAergic.} strength of the synaptic connection from ACell $\Cell{A}{j}$ to BCell $\Cell{B}{i}$ with the convention that $\W{A}{j}{B}{i}=0$ if there is no connection from $\Cell{A}{j}$ to $\Cell{B}{i}$. In general $\W{A}{j}{B}{i} \neq \W{A}{i}{B}{j}$.
 Likewise, 
BCells are connected to ACells via a connectivity matrix $\W{B}{}{A}{}$ where the synapses $\W{B}{i}{A}{j} \geq 0$ (excitatory).

We consider synaptic responses modelled by exponential alpha profiles \cite{destexhe-mainen-etal:94b}. Thus, the PSP induced by ACells on BCell $i$ is: 
\begin{equation}\label{eq:PSP_A_Bip}
\Pst{B}{i}(t) = \sum_{j=1}^{N_B} \W{A}{j}{B}{i} \int_{-\infty}^t \alpha_{B}(t-s) \, \V{A}{j}(s) ds,  \qquad \alpha_{B}(t)=e^{-\frac{t}{\tau_{B}}} H(t),
\end{equation}
where the Heaviside function $H$ ensures causality.

Conversely, the BCell $\Cell{B}{i}$ connected to $\Cell{A}{j}$  induces, on this cell, a synaptic response characterized by a post-synaptic potential (PSP) $\Pst{A}{j}(t)$. We consider here passive ACells so that their voltage $\V{A}{j}(t)$ is equal to this PSP:
\begin{equation}\label{eq:PSP_Bip_A}
\V{A}{j}(t) = \sum_{i=1}^{N_A} \W{B}{i}{A}{j} \int_{-\infty}^t \alpha_{A}(t-s) \, \R{B}{i}(s) ds,
\end{equation}
with $\alpha_{A}(t)=e^{-\frac{t}{\tau_{A}}} H(t)$. Note that the voltage of the BCell is rectified and gain-controlled.

\subsection{Waves of activity induced by a stimulus}\label{Sec:WaveStimulus}

The question we are interested in is the following. Consider a stimulus moving through the receptive field of BCells. How do ACells lateral connectivity  influences the response of BCells to that stimulus ? To answer to this question we rewrite the joint dynamics of BCells voltage, BCells activity, ACells voltage in the form of a $3N$ dimensional dynamical system. We use Greek indices $\alpha,\beta,\gamma = 1 \dots 3N$ and define the state vector $\cX$ as  
$$\vcX_\alpha =
\left\{
\begin{array}{llll}
&\V{B}{i}, \quad \alpha=i; \\
&\V{A}{i}, \quad \alpha=N+i; \\
&A_i, \quad \alpha=2N+i;
\end{array}
\quad i=1 \dots N.
\right.
$$
Likewise, we define the stimulus vector $\vcF_\alpha=\F{B}{i}$, if $\alpha=1 \dots N$ and  $\vcF_\alpha=0$ otherwise. $\F{B}{i}$ is a function of the drive \eqref{eq:Vdrive}.
Then, the dynamical system describing the joint dynamics has the general form:
\begin{equation}\label{eq:Diff_Syst_Vect_Gen}
\frac{d \vcX}{dt} = \cH(\vcX)  + \vcF(t).
\end{equation}
where $\cH(\vcX)$ is a non linear function, due to function $\R{B}{i}\pare{\V{B}{i}, \A{B}{i}}$, featuring gain control and low voltage threshold. This system is a non linear, non autonomous, dynamical system. 

We don't know about any general method to analyse this dynamics.
However, the non linearity comes from the function\eqref{eq:gain_control_bip}, $\cG_B(\A{B}{i})$ which can be approximated by a piecewise linear function.  
The idea of using such a phase space decomposition with piecewise linear approximations has been used, in a different context by S. Coombes et al \cite{coombes-lai-etal:18} and in \cite{cessac:08,cessac-vieville:08,cessac:11}. Then, there is a domain $\Omega$ of $\setR^{3N}$, where $\R{B}{i}\pare{\V{B}{i}, \A{B}{i}}=\V{B}{i}$ so that \eqref{eq:Diff_Syst_Vect_Gen} is linear and can be written in the form :
\begin{equation}\label{eq:Diff_Syst_Vect_Lin}
\frac{d \vcX}{dt} = \cL.\vcX  + \vcF(t).
\end{equation}
where $\cL$ is a $3N \times 3N$ matrix, depending on the connectivity matrices $\W{A}{}{B}{},\W{B}{}{A}{}$ (see \cite{souihel-cessac:20} for detail). Being in $\Omega$ corresponds to intermediate activity, where neither BCells gain control \eqref{eq:gain_control_bip} nor low threshold \eqref{eq:NLinRectBip} are active. One can  then study this case and describe then what happens when trajectories of \eqref{eq:Diff_Syst_Vect_Gen} get out of this domain, activating low voltage threshold or gain control.

The general solution of \eqref{eq:Diff_Syst_Vect_Lin} is:
\begin{equation}\label{eq:GenSolSDLin}
\vcX(t)=\int_{t_0}^t e^{\cL(t-s)}.\vcF(s) \, ds,
\end{equation}
The behaviour of the solution \eqref{eq:GenSolSDLin} depends on the eigenvalues $\lambda_\beta, \beta=1 \dots 3N$ of $\cL$ and its eigenvectors, $\vcP_\beta$, with entries $\cP_{\alpha\beta}$. The matrix $\cP$ transforms $\cL$ in Jordan form ($\cL$ is not diagonalizable when $h_B \neq 0$). Whatever the form of the connectivity matrices $\W{B}{}{A}{},\W{A}{}{B}{}$ the $N$ last eigenvalues are always $\lambda_\beta=-\frac{1}{\tau_a}, \beta=2N+1 \dots 3N$, where $\tau_a$ is the characteristic time of the BCells activity variable \eqref{eq:dA}.\\

From \eqref{eq:GenSolSDLin}, one can show the following general result (not depending on the specific form of $\W{B}{}{A}{},\W{A}{}{B}{}$, they just need to be square matrices and to be diagonalizable):
\begin{equation}\label{eq:XalphaDriven}
\cX_\alpha(t) = V_{\alpha_{drive}}(t) \, + \E{B}{B,\alpha}(t)+\E{B}{A,\alpha}(t)+\E{B}{a,\alpha}(t), \quad \alpha =1 \dots 3N,
\end{equation}
where the drive term \eqref{eq:Vdrive} is extended here to $3N$-dimensions with $V_{\alpha_{drive}}(t)=0$ if $\alpha > N$. It corresponds to the direct effect of the drive on BCells and this is the only non vanishing term when $\W{B}{}{A}{}=\W{A}{}{B}{}=0$. 
The other terms have the following definition and meaning. $\E{B}{B,\alpha}(t)$,  corresponds to the \textit{indirect} effect, via the ACells connectivity, of the BCells drive on BCells voltages (i.e. the drive excites BCell $i$ which acts on BCell $j$ via the ACells network);
$\E{B}{A,\alpha}(t)$ corresponds to the effect of BCell drive on ACells voltages, and, finally $\E{B}{a,\alpha}(t)$
corresponds to the effect of the BCells drive on the dynamics of BCell activity variables (see \cite{souihel-cessac:20} for the explicit form).  All these terms involves the spectrum of $\cL$ and its eigenvectors, making a connection between spectral properties and the network response to a spatio-temporal stimulus. 

In particular, the eigenvalues of $\cL$ can be real or complex, stable (real part $<0$)  or unstable, essentially depending on the connectivity matrices. In this linear response, the network effect is the superimposition of stable or unstable eigenmodes that can interfere in a constructive way, amplifying the stimulus response, and generate a wave of activity propagating ahead of the stimulus \cite{souihel-cessac:20}. Depending on the connectivity matrices (graph structure and intensity of weights) these eigenmodes can also interfere and weaken the response to stimuli. Depending on parameters too, the dynamics \eqref{eq:Diff_Syst_Vect_Gen} can stay within the set $\Omega$ where the linear approximation \eqref{eq:Diff_Syst_Vect_Lin} holds. However, the interesting regime is when dynamics gets out of $\Omega$. It happens if some eigenmodes are unstable, but it can happen too when all modes stable, whenever the cumulative of the drive and network effect decreases the BCell voltage sufficiently low to activate the rectification term, or increases the BCell activity sufficiently to trigger gain control \eqref{eq:gain_control_bip}. In all cases getting out of $\Omega$ generates a mechanism saturating the instability, preventing divergence and thereby stabilizing the cell in a range where it is still able to respond to the stimuli. In particular, the activation of the gain control mechanism induces anticipation and this activation can be enhanced by the wave induced from the ACells network.

\subsection{Consequences}\label{Sec:ConsequencesAnticipation}

It is known that the visual system is able to anticipate the motion of an object moving on a deterministic, regular, trajectory \cite{valois-valois:91,nijhawan:94,baldo-klein:95,nijhawan:97}. 
In the early visual cortex an object moving across the visual field triggers a wave of activity ahead of motion, thanks to the cortical lateral connectivity \cite{benvenuti-chemla-etal:20,subramaniyan-ecker-etal:18,jancke-erlaghen-etal:04}. 
In the retina, anticipation takes a different form. One observes a peak in the firing rate response of GCells to a moving object, occurring \textit{before} the peak response to the same object when flashed. This effect can be explained by purely local mechanisms, at individual cells level, like gain control, here characterized by \eqref{eq:gain_control_bip} \cite{berry-brivanlou-etal:99,chen-marre-etal:13}. To our best knowledge, collective effects similar to the cortical ones - that is, a rise in the cell’s activity before the object enters in its receptive field due to a wave of activity ahead of the moving object - have not been reported yet experimentally. 

In contrast, our analysis predicts such an effect and emphasizes a potential role of amacrine lateral connectivity on anticipation (gap junctions may also play a role but we don't comment it here). The connectivity graph is involved in the spectrum of a propagation operator controlling the time evolution of the network response to a moving stimulus. In the linear approximation, we obtain a superimposition of several effects which can collaborate and enhance anticipation, or compete. \\%Non linearity, here, is local and can be treated in the sa	me %Thisgraph can be instantiated by studying differential motion sensitive cells \cite{baccus-meister:02} (i.e. cells which are able to respond to an object moving upon a moving background, with a different motion)  with two types of connectivity: nearest neighbours, and a random connectivity, inspired from biology \cite{tauchi-masland:84}.

%We actually observe two forms of anticipation. The first one, discussed in the beginning of this introduction and already observed in \cite{berry-brivanlou-etal:99,chen-marre-etal:13}, is a shift in the peak of a retinal Gcell response, occurring before the object reaches the center of its receptive field. In our case, lateral connectivity can enhance the shift improving the mere effect of gain control. The second anticipation effect we observe is a raise in GCells activity before the bar reaches the receptive field of the cell, similarly to what is observed in the cortex \cite{benvenuti-chemla-etal:20}. To the best of our knowledge, this effect has not been studied in the retina and constitutes therefore a prediction of our model.\\

%In this section we have mathematically argued that %a moving stimulus can - under specific conditions mathematically controlled - induce a wave of activity which propagates ahead of the stimulus thanks to lateral connectivity. This suggests that,
% in addition to local gain control mechanism inducing an anticipated peak of GCells activity, lateral connectivity could induce a mechanism of neural latencies reduction, similar to what is observed in the cortex \cite{benvenuti-chemla-etal:20,subramaniyan-ecker-etal:18,jancke-erlaghen-etal:04}. 

Our model was constructed with a simplification of ACells dynamics, which are simple linear integrators. Non-linear mechanisms in the dynamics of these cells could enhance resonance-like effects in the network and, thereby, favour the propagation of a lateral wave of activity induced by a moving stimulus \cite{souihel-cessac:20}. Thus, a more detailed analysis would require a closer investigation of BCells and ACells dynamics and connectivity, especially implying to define more specifically the type of ACell (AII, Starburst, A17, wide field, medium field, narrow field, ...) and the type of functional circuit one wants to consider.% Note that ACells are difficult to access experimentally due to their location inside the retina.

The retina is only the early stage of the visual system. Visual responses are then processed via the thalamus and the cortex. As exposed in the introduction, anticipation is also observed in V1 with a different modality than in the retina. How do these two effects combine ? How does retinal
anticipation impact cortical anticipation ? To answer these questions at a computational level one would need to propose a model of the retino-thalamico-cortical pathway which, to the best of our knowledge, has never been done. This a work in progress (S. Souihel, M. di Volo, S. Chemla, A. Destexhe, F. Chavane and B. Cessac., in preparation). See \cite{souihel:19} for preliminary results.

\section{The structure of correlations}\label{Sec:Correlations}

The trajectory of an object - which is, in general, quite more complex than a moving bar with constant speed - across our visual field generally involves long-range correlations in space and in time. Local information about this motion is encoded by retinal ganglion cells. Decoders based on the firing rates of these cells can extract some of the motion features \cite{palmer-marre-etal:15,salisbury-palmer:16,deny-ferrari-etal:17,sederberg-maclean-etal:18,srinivasan-laughlin-etal:82,hosoya-baccus-etal:05,kastner-baccus:13}. Yet, lateral connectivity in the retina - especially via amacrine cells connecting bipolar cells - plays a central role in motion processing (see e.g.  \cite{gollisch-meister:10}). In addition, ganglion cells are directly connected through electric synapses and indirectly via ACells.  What could be the role of this lateral connectivity in motion processing ? Clearly, one may expect it to induce spatial and temporal correlations in spiking activity, as an echo, a trace, of the object's  trajectory. These correlations cannot be read in the variations of firing rate; they also cannot be read in synchronous pairwise correlations as the propagation of information due to lateral connectivity necessarily involves \textit{delays}. This example raises the question about what information can be extracted from spatio-temporal correlations in a 
network of connected neurons submitted to a transient stimulus. 
What is the effect of the stimulus on these correlations ? How can one handle this information from data where one has to measure \textit{transient} correlations ? 

Here, we address these questions in a context linking dynamics to statistics of orbits, that is, ergodic theory and statistical mechanics.
 
\subsection{The spontaneous activity of Ganglion cells}\label{Sec:SponGC}

GCells are spiking cells. Henceforth, they can be described either by a spiking neuronal model, like the Integrate and Fire \cite{lapicque:07}, or by a firing rate model. We consider a dual approach where GCells spike according to a joint firing probability of the form:
\begin{equation}\label{eq:PtransGCs}
\Pnc{\omega(n)}{\sif{n-1}} \equiv e^{\phi(n,\omega)},
\end{equation}
Here, time is binned with a discrete time interval $\delta>0$ (typically a few ms). We consider a set of $N$ GCells with index $k=1 \dots N$. We are just interested in the "spike state" of GCells, defined by a variable $\omega_k(n) \in \Set{0,1}$: $\omega_k(n)=1$ if GCell $k$ spikes in the time bin $n$ and $\omega_k(n)=0$ otherwise. The vector of all spike states ("spike pattern") at time bin $n$ is noted $\omega(n)=\vect{\omega_k(n)}_{k=1}^N$. A "spike history" is a list of successive spike patterns from some time $m$ to some time $n \geq m$, $\bloc{m}{n} \equiv \omega(m),\omega(m+1), \dots, \omega(n)$. Thus, $\sif{n-1}$ denotes the history prior to time $n$, where the past goes to $-\infty$. Therefore, $\Pnc{\omega(n)}{\sif{n-1}}$ is the conditional probability to observe the spike pattern $n$ given the infinite history prior to $n$. The index $n$, in $\Pnc{}{}$, expresses that the transition probabilities depend explicity on time (we don't assume stationarity). Note that, such transition probabilities, generalizing the notion of Markov chain to infinite memory are rigorously defined provided they decay sufficiently fast (e.g. exponentially) with the history depth. Under suitable conditions
\footnote{We assume that $\Pnc{\omega(n)}{\sif{n-1}}>0$ so that $\phi(n,\omega) = \log \Pnc{\omega(n)}{\sif{n-1}} > - \infty$.
We note, for $n \in \mathds{Z}$, $m \geq 0$, and $r$ integer:
$$\omega \stackrel{m,n}{=} \omega', \,  \mbox{ if } \omega(r) = \omega'(r), \forall r \in \{n-m, . . . , n\}.$$
For $m$ be a positive integer, the $m$-variation of  $\Pnc{\omega(n)}{\cdot}$ is:
$$
var_m[\Pnc{\omega(n)}{\cdot}]=\sup\left\lbrace \mid \Pnc{\omega(n)}{\sif{n-1}}-\mathds{P}_n[\omega(n)\mid \omega'^{n-1}_{-\infty}] \mid : \omega \stackrel{m,n}{=} \omega'\right\rbrace.
$$
A sufficient condition to ensure that this system of transition probabilities has a unique invariant measure compatible with it is that $var_m$ decays exponentially fast with $m$ \cite{fernandez-maillard:05}.
}
there is a unique invariant probability associated with this set of transition probabilities (generalizing the unique invariant probability of a Markov chain with a primitive transition matrix \cite{gantmacher:98,seneta:06}), called "Chain with complete connection" \cite{onicescu-mihoc:35,fernandez-maillard:05} in probability theory, "g-measure"  in ergodic theory \cite{keane:70}, or "Gibbs measure". For the last terminology they have indeed a close connection to their statistical mechanics counterpart \cite{georgii:88,leny:08,cofre-maldonado-etal:20}, if one assimilates the sequence of spike patterns as a left-sided infinite chain of sites, labelled by $n$, the time index, with states taking values in $\Set{0,1}^N$. These sites can then be assimilated to "spins" taking $2^N$ possible values. In eq. \eqref{eq:PtransGCs},
 $\phi(n,\omega)$ is then naturally called a (normalised) Gibbs potential. It is normalised because it is the log of a probability so there is no need to divide by a partition function (see \cite{cofre-cessac:14,cofre-maldonado-etal:20} for the link between "thermodynamic" like Gibbs potentials and normalized potentials).\\
  
What is the link of this rather abstract paragraph and GCells in the retina ? The idea is the following. It is experimentally easy to measure GCells collective activity (voltage, from Multi-Electrodes Arrays) and to extract spikes via spike sorting \cite{yger-spampinato-etal:18,buccino-hurwitz-etal:19}. From this information one attempts to construct a probability distribution that mimics the collective spikes statistics. This statistics is constrained by "hidden" information, typically the network of amacrine cells which correlates GCells by indirect interactions, or gap junctions which connect them directly. Of course, the applied stimulus plays a role and what the GCells receive is the processing of this stimulus through the whole retinal network above GCells (see section \ref{Sec:Retina}). If one has only information about the spikes the most general form of statistics is \eqref{eq:PtransGCs}. The probability to observe a spike pattern in a time bin $n$ depends on the whole spike history of the GCells network. This history includes the stimulus influence. Of course, for specificity and tractability, one has to consider simplifications of this formalism, for example that successive times are independent.
In this case, one obtains a class of models usually designed as Maximum Entropy models (although Maximum Entropy principle is quite more general than that \cite{cessac-cofre:13}).
A classical model in this setting is the so-called Ising model - where the potential $\phi$
takes the form of the energy in a spin-glass with binary spins - or variations (type dependent local field, triplets or quadruplets interactions) \cite{schneidman-berry-etal:06,shlens-field-etal:06,tkacik-prentice-etal:10,ganmor-segev-etal:11,ganmor-segev-etal:11b,tkacik-marre-etal:13b,granot-atedgi:13,nghiem-telenczuk-etal:18}, see \cite{gardella-marre-etal:19} for a physicists-oriented review.

Another approach, considering, causality and time correlations, consists of considering conditional probabilities of the form:
\begin{equation}\label{eq:GLM}
\Pnc{\omega_k(n)=1}{\sif{n-1}} =f\pare{\sum_{j \in \cB_k}  \sum_{l \leq n-1} J_{kj}(l)\, \omega_j(l) +  S_k(n)}
\end{equation}
called Generalized Linear Model \cite{pillow-shlens-etal:08}. $f$ is a non linear function. As in section \ref{Sec:Waves}, $\cB_k$ is a set of cells connected to neuron $k$, $S_k(n)$ is the stimulus viewed by neuron $k$, namely processed through the retina network. The $J_{kj}(l)$s corresponds to "effective" interactions. In a classical neuronal model this would be the direct synaptic interactions between neurons, with a possible time integration reflected in the time dependence (see section \ref{Sec:BMS} for an example). Here, GCells are not directly connected so the $J_{kj}(l)$s should be considered as parameters that need to be fit to best approximate statistics. 

MaxEnt and GLM are in fact closely related through Gibbs distributions as developed in 
\cite{cessac-cofre:13,cofre-maldonado-etal:20}.

\subsection{An example: the discrete time integrate and fire model}\label{Sec:BMS}

There are many models for neuronal dynamics both at the level of individual neurons and neuronal networks \cite{dayan-abbott:01,gerstner-kistler:02,ermentrout-terman:10}. A canonical example of such a model, actually the first one proposed to the scientific community was introduced by Lapicque in 1907 \cite{lapicque:07}, the Integrate and Fire model. Here, we consider a discrete time version of this model (and discuss the extension to continuous time at the end of the section). We consider $N$ neurons with index $k=1 \dots N$ and membrane potential $V_k$, depending on a discrete time $n$. We fix a variable $\theta$ called "spiking threshold" such that voltages dynamics obeys \cite{cessac:08,cessac:11}:
\begin{equation}\label{eq:BMS}
\left\{
\begin{array}{lll}
V^k(n+1)&=\gamma \, V^k(n) + \sum_{j \in \cB(k)} W_{kj} \omega_j(n)  + I_k(n) + \sigma_B \xi_k (n), \quad &\mbox{if } V^k(n)< \theta, \quad \mbox{Integrate\,phase};\\
&\\
V^k(n+1)&=0 \quad \mbox{and} \quad \omega_k(n)=1, \quad &\mbox{if } V^k(n) \geq \theta, \quad \mbox{Firing\,phase},
\end{array}
\right.
\end{equation}
where $0 < \gamma < 1$. If, at some, discrete time $n$, $V^k(n)$ exceeds the threshold $\theta$ the membrane potential is reset at time $n+1$ and a spike is recorded, at $n$ for neuron $k$, i.e. $\omega_k(n)=1$. We have assumed that synapses are instantaneous.  Then, the synaptic input is $\sum_j W_{kj} \omega_j(n)$, that is a pre-synaptic neuron $j$ acts on the post-synaptic neuron $k$ whenever $j$ spikes, $\omega_j(n)=1$. $I_k(n)$ is the time discretisation of the external stimulus influence (for example, a time discretisation of the BCell response \eqref{eq:Vdrive}). Note that it depends on $n$, hence dynamics \eqref{eq:BMS} is non stationary.  $\xi_k (n)$ are  independent standard Gaussian random variables representing the effect of noise.

It is easy to integrate equation \eqref{eq:BMS} conditionally upon a fixed spike sequence $\omega$. A trajectory $V \equiv \Set{V_k(n), k=1 \dots N, n \in \setZ}$ is \textit{compatible}\footnote{Compatibility has to do here with symbolic coding of voltage trajectories and Markov partitions, see \cite{cessac:08} for detail.} with the spike sequence $\omega$ if $\chi\pare{V_k(n) > \theta} = \omega_k(n)$, $\forall k=1 \dots N, n \in \setZ$, where $\chi\pare{A}$ is the indicator function of the logical event $A$, $\chi\pare{A}=1$ if $A$ is true, $\chi\pare{A}=0$ otherwise.  We note $\nko=\max \Set{l, l < n \, | \, \omega_k(n)=1}$ the last time before $n$ where neuron $k$ has spiked, thus whose voltage was reset to $0$. Then:
\begin{equation}\label{eq:IntegBMS}
V^k(n+1)=\sum_{j=1}^N W_{kj} \, \eta_{kj}(n,x) + \sum_{l=\nko}^n \gamma^{n-l} I_k(l)+ \sigma_B \sum_{l=\nko}^n \gamma^{n-l} \xi_k(l) ,
\end{equation} 
where:
\begin{equation}\label{eq:eta}
 \eta_{kj}(n,\omega)=\sum_{l=\nko}^n \gamma^{n-l} \, \omega_j(l),
\end{equation}
integrates the influence of pre-synaptic neuron $j$ on the time interval $\bra{\nko+1,n}$. Each spike emitted by this neuron, at time $l$ in this time interval, contributes with the weight $\gamma^{n-l}$ and there is no contribution at times where this neuron didn't spike, $\omega_j(l)=0$.
The condition $\gamma < 1$ somewhat defines the exponential decay in the spike history dependence via the characteristic time:
\begin{equation}\label{eq:tau_gamma}
\tau_\gamma=-\ent{\frac{1}{\log\gamma}}.
\end{equation}
Likewise, $\sum_{l=\nko}^n \gamma^{n-l} I_k(l)$ integrates the stimulus influence on neuron $k$ and $\sigma_B \sum_{l=\nko}^n \gamma^{n-l} \xi_k(l)$ is the integrated noise term. This is a Gaussian random variable, with mean zero and variance $\sigma(k,n,x) = \sigma_B^2 \, \frac{1 - \gamma^{2(n+1-\nko)}}{1-\gamma^2}$.\\

Thanks to the integrated equation \eqref{eq:IntegBMS} and because the integrated noise is Gaussian it is now easy to compute the probability that neuron $k$ spikes at time $n+1$ \textit{given} the history $\omega$, $\Pnc{\omega_k(n)=1}{\sif{n-1}} = \Pnc{V^k_{n} \geq \theta}{\sif{n-1}}$. This is:
\begin{equation}\label{eq:PMargBMS}
\Pnc{\omega_k(n)=1}{\sif{n-1}} = f\pare{\frac{\theta - \sum_{j=1}^N W_{kj} \, \eta_{kj}(n,\omega)  - \sum_{l=\nko}^n \gamma^{n-l} I_k(l) }{\sigma_B \, \sqrt{\frac{1 - \gamma^{2(n+1-\nko)}}{1-\gamma^2}}}}
\end{equation}
where $f(z)=\int_z^{+\infty} \, \frac{e^{-\frac{u^2}{2}}}{\sqrt{2 \pi}} \, dz$.

 As integrated noise variables for different $k$s are independent we have, for any spike pattern $\omega(n)$:
\begin{equation}\label{eq:PJoinBMS}
\Pnc{\omega(n)}{\sif{n-1}}  = \prod_{k=1}^N  \pare{\omega_k(n)} \, \Pnc{\omega_k(n)=1}{\sif{n-1}}  + \pare{1-\omega_k(n)} \, \pare{1-\Pnc{\omega_k(n)=1}{\sif{n-1}}},
\end{equation}
giving an explicit form to \eqref{eq:PtransGCs} and linking explicitly the dynamics \eqref{eq:BMS} to statistics, with an explicit dependence in network parameters. Note, in particular, the form similar to the GLM model \eqref{eq:GLM}, although quite more general (as the $J_{ij}(l)=W_{ij} \gamma^{n-l}$ are here explicitly written). 

Let us comment this result. First, \eqref{eq:PJoinBMS} is the transition probability, of the form \eqref{eq:PtransGCs}  where the normalised Gibbs potential $\phi$ can be explicitly written, 
in terms of the synaptic interactions and the parameter $\sigma_B$ controlling the noise. Here, the memory depth, going back to $\nko$, the last time in the past where neuron $k$ was reset, depends on $\omega$, defining a  \textit{variable length Markov chain}. In addition, $\nko$, has a function of $\omega$, is unbounded. One can find for all $r \in \Set{-\infty,n}$ a sequence $\omega$ such that $\nko=r$ (take the sequences where all $\omega_k(n)=0$, $k=1 \dots N$, $n > r$ and $\omega_k(n)=1$ for at least one $k$). We have thus to deal with the extension of Markov chains, to chains with unbounded memory also called chains with complete connections \cite{onicescu-mihoc:35}. The existence of a Gibbs distribution compatible with this chain is guarantee by the exponential decay of the memory controlled by $\gamma <1$ \cite{cessac:11b}. Finally, in \eqref{eq:PJoinBMS}, the transition probabilities depend explicitly  on time, because of the stimulus dependent term. They are therefore  non stationary.

The form \eqref{eq:GLM} can also  be mathematically justified for a network of Integrate and Fire neurons with continuous time \cite{cofre-cessac:13}. The main difficulty is that, in this case, the set of spikes in uncountable (because the spike is instantaneous in IF models, which is a biological, physical, and mathematical non sense). Using the realistic fact that spikes have a duration it is nevertheless possible to associate to the continuous time dynamics of the IF a countable set of spikes \cite{cessac-vieville:08}, for which one can establishes a generalisation of eq. \eqref{eq:PMargBMS}, which, actually, allows to consider conductance-based IF \cite{rudolph-destexhe:06} where the conductance depends on the spike history. The form \eqref{eq:GLM}  fails when there are fast synapses such as gap junctions and more general cases have to be considered \cite{cofre-cessac:13}, which encompass Maximum Entropy models and GLMs in the formalism of Gibbs distributions \cite{cessac-cofre:13,cofre-cessac:14}.

\subsection{Response to stimuli}\label{Sec:ResponseStimuli}

The main advantage of the general form \eqref{eq:PtransGCs} is that it handles non stationarity, i.e. the dependence of the statistics in a spatio-temporal stimulus.
More explicitly, a form like \eqref{eq:PtransGCs} shows that the statistics of spikes at time $n$ depends on a delicate alchemy involving intrinsic neurons dynamics, their interactions, and the stimulus. 

Obviously, the form \eqref{eq:PtransGCs} is too general to be useful unless additional assumptions are made. Such a natural assumption is to consider that the stimulus is "small" and additive, that is, that one can make a Taylor expansion of the potential as powers of the stimulus. In this case, it is possible to show that the response of the neuronal system to a stimulus, that is, the variations in spike statistics induced by the stimulation, can be described in terms of a linear response theory \cite{cessac-ampuero-etal:20}.

The main result is that the variation, in the average of an observable $f$ (i.e. a real function of spikes blocs), resulting from the application of a stimulus reads:
\begin{equation} \label{eq:LinRep}
\delta \mu\bra{f}(n) \equiv \mu^S\bra{f}(n) - \mu^{(sp)}\bra{f} = \pare{K_f \ast S}(n),
\end{equation}  
where $\mu^{(sp)}$ is the Gibbs distribution in spontaneous activity (without stimulus), $\mu^S$ is the Gibbs distribution in evoked activity (with stimulus), $\mu^S\bra{f}(n)$ means the average of $f$ with respect to $\mu^S$, which depends on time (if the stimulus does), and $\mu^{(sp)}\bra{f}$ means the average of $f$ with respect to $\mu^{(sp)}$, which does not depends on time. This variation in average is given by a convolution of a kernel $K_f$, depending on $f$, and of the stimulus. 

The form \eqref{eq:LinRep} is classical, looking very much like, e.g., the RF modelling \eqref{eq:Vdrive}. The detail of this simple canonical form hides in fact a high complexity. In particular, it depends on $f$ and on the parameters determining dynamics (e.g. synaptic weights). In general, $K_f$ can be expressed as a sum  correlations between $f$ and spikes events of the form $\omega_{k_1}(n_1) \dots \omega_{k_r}(n_r)$, called monomials or order $r$. Monomials of order $1$ have the form $\omega_{k_1}(n_1)$ and just tell us if neuron $k_1$ is spiking at time $n_1$, monomials of order $2$ have the form $\omega_{k_1}(n_1)\omega_{k_2}(n_2)$ and tell us if neuron $k_1$ is spiking at time $n_1$ \textit{and} neuron $k_2$ is spiking at time $n_2$, and so on. The correlations are computed with respect to the stationary measure $\mu^{(sp)}$ so that only the difference in times matters when computing correlations. In addition, the influence of monomials in the expansion decreases with their order, so that one can obtain a reasonable expression of the convolution kernel considering averages of order two monomials (time-dependent pairwise correlations).

This is therefore a result in the form of a fluctuation-dissipation "theorem" in statistical physics, with the difference that one considers time dependent correlations. This formula has proven to give astonishingly good results when computing the response to a time dependent stimulus in the moel \eqref{eq:BMS} \cite{cessac-ampuero-etal:20}.

\subsection{Consequences}\label{Sec:ConsequencesCorr}

The result presented here introduces a general mathematical formalism allowing one to handle spike correlations as a result of a neuronal network activity in response to a stimulus. The most salient result of this work is that the difference of an observable average in response to an time dependent external stimulus of weak amplitude can be computed from the knowledge of the spontaneous correlations, i.e., from the dynamics without the stimulus. This result is not surprising from a non-equilibrium statistical physics perspective (Kubo relations, fluctuation-dissipation relation \cite{kubo:66, ruelle:99}). However, to the best of our knowledge, it has been established for the first time for non stationary spiking neuronal networks in the recent paper \cite{cessac-ampuero-etal:20}. The novelty of our approach is that it provides a consistent treatment of the expected perturbation of higher-order correlations, going in this way, beyond the known linear perturbation of firing rates and instantaneous pairwise correlations; in particular, it extends to time-dependent correlations.

As we have shown, and as expected, the stimulus-response and dynamics are entangled in a complex manner. For example, the response of a neuron $k$ to a stimulus applied on neuron $j$ does not only depends on the synaptic weight $W_{kj}$ but, in general, on all synaptic weights, because the dynamics create complex causality loops which build up the response of neuron $k$ \cite{cofre-cessac:14, cessac-sepulchre:04,cessac-sepulchre:06}. We formally obtained a linear response function in terms of the parameters of a spiking neuronal network model and the spike history of the network. 
Although a linear treatment may seem a strong simplification, our results suggest that already, in this case, the connectivity architecture should not be neglected. In the presence of stimuli, the whole architecture of synaptic connectivity, history and the dynamical properties of the networks are playing a role in the correlations through the perturbed potential. It follows from our analysis that this effective connectivity depends on stimulus, as already observed for the effective connectivity defined through the Ising model \cite{cocco-leibler-etal:09}; it depends also on synaptic weights, in a complex manner. It was observed in \cite{cessac-sepulchre:06} that the existence of resonances in the power spectrum generates stimuli dependent graphs e.g. controlled by
the frequency of an harmonic stimulus. It would be interesting to check whether such a property holds here as well.

\section{Discussion and perspectives}\label{Sec:Discussion}

Although the retina might be viewed as nothing more than a part of the brain, it shows up strong differences, which has to be thoroughly considered for a modeller interest in the retina dynamics. In this paper, we have shown how specific behaviours of the retinal network can be considered in the context of dynamical systems theory. Yet, a large part of the modelling and of the mathematical machinery used for our models analysis are not really specific to the retina and could be addressed in the cortex as well.

In this section we would like to consider questions, more specific to the retina structure and dynamics, that could be tackled in the realm of dynamical systems theory. 

\subsection{Functional retinal circuits}\label{Sec:RetinalCircuits}

The feed-forward / lateral structure of the retina, the diversity of cells subtypes for a given type (eg, around 40 amacrine cell subtypes), synaptic types offer a gigantic combinatorial of potential retinal circuits. There are however few functional circuits, duly identified, and characterized from the experimental point of view. An important example are the ON, OFF, and ON-OFF pathways \cite{nelson-kolb:04}. A central concept in visual system study is the notion of “feature detectors” where the visual system separates the visual image into components. From this perspective highlights and shadows are two of the most fundamental features of images. In all vertebrate visual systems the separation of highlights and shadows is performed
by two separate pathways, starting at the retina : the ON and the OFF pathways. This organization allows to provide information to the ganglion cells concerning brighter than background stimuli (ON-center) or darker than background stimuli (OFF-center). More generally,
the trajectory of an object (e.g. a bright bar) on a
moving background necessarily involves these
pathways performing local detection of highlights
and shadows. These two pathways are, in fact, connected via AII amacrine cells and gap
junctions, indirectly connecting ganglion cells thereby potentially creating spatio-temporal correlations between local information channels.  This example outlines the functional role of specific retina circuits in processing visual information.
Other pathways implied in differential motion would also be interesting to study, as discussed in section \ref{Sec:Response}. 

These circuits are small in the sense that they involve a few, specific cells, with specific synapses. Here, thus, the details of the biophysical modelling play a distinguished role, especially the physiological parameters involved (see also section \ref{Sec:DiscParameters}). 
Yet, these circuits performing local information extraction from a visual scene, span the whole retina, resulting in a multi-scale, stimulus-driven dynamics, that can hardly be treated with classical mean-field methods used in the cortex  \cite{faugeras-touboul-etal:09,schuecker-goedeke-etal:16,helias-dahmen:20}, especially taking into account that most of the involved neurons are not spiking. A study in this spirit has been proposed in the section \ref{Sec:Waves} dealing with retinal waves. Yet, the situation was still "simple"
as we were considering one type of cells and synapses, without sensory entry. A more general, multi-scale, study of retina circuits like the ON-OFF would deserve considerably more efforts, but, is, perhaps, the necessary way to understand how retina efficiently encodes a visual scene.  

\subsection{Changing physiological parameters}\label{Sec:DiscParameters}

The interest of a careful modelling of the retina by dynamical systems is also found in the impact of variations of parameters on local and global functioning of the retina. 
How do specific local characteristics of neurons, synapses, retinal circuits - which may be affected by physiological modifications (development, short term plasticity, pharmacology, pathology) - impact the large-scale dynamics of the retina, in particular in the presence of spatio-temporal visual stimuli ? Living systems, such as the retina, are somewhat robust to small variations in their physiological parameters ... up to some extent. One of the profound lessons to be drawn from the theory of bifurcations is that continuous, smooth, variations of parameters lead to smooth changes ... until bifurcations points are reached. There, changes are dramatic and drastic. The interest of studying retinal circuits, from the local scale to the whole retina scale, in a dynamical systems perspective, is precisely to anticipate potential bifurcations induced by physiological parameters variations. An example as been presented in section \ref{Sec:Waves}. In the related paper \cite{matzakou-karvouniari-gil-etal:19} the effect of variations of parameters such as the conductance of fast potassium channels has lead us to conjecture a role played by specific channels, called Kv3, and the impact of pharmacological actions. In this perspective, dynamical systems modelling and bifurcations theory are theoretical complement to experimental investigations aiming at better understand the functioning of the retina in normal ... or pathological conditions.

Let us indeed briefly deal with retinal pathologies. Visual impairment affects millions of people in the world.
About $39$ million were considered blind in 2014 \cite{lok:14}.
Roughly $80\%$ of visual impairment is preventable or curable, including operable conditions such as cataracts. However, retinal-degeneration disorders 
had no cure until recently \cite{lok:14}. Retinal degeneration is the deterioration of the retina caused by the progressive and eventual death of the cells of the retina. Important examples are Age Macular Degeneration (AMD) or Retinitis Pigmentosa .

In the recent years, a new hope of slowing or even reversing vision
loss from retinal disorders has grown with technics such as gene therapy, cell transplants and retinal prostheses which have reached the level of clinical trials testing, mostly in treating
rare congenital disorders.  In parallel, several biotechnology firms have formed to take treatments through clinical testing. Institutes, like the Institut de la Vision in Paris, 
at the intersection between fundamental and clinical research, and industrial development are born. There are large avenues for thorough modelling of these diseases, the progressive degenerations of cells, the effect of therapies, using methods from dynamical systems theory. A promising example is the modelling of the spread and treatment of hyperoxia-induced photoreceptor degeneration in retinitis pigmentosa  \cite{camacho-punzo-etal:16,roberts-gaffney-etal:17,roberts-gaffney-etal:18}. Such an approach could also benefit to the large development of retinal prostheses, technological devices  used to stimulate specific, still responsive, cells in a degenerated retina, thereby (partially) restoring vision. For example, a modelling approach combined with dynamical system approach could be used to characterize the effect and the optimal regime of electric stimulation with electrodes, in a retina whose physiology, circuitry, and dynamics constantly evolve along the progression of the disease \cite{marc-jones-etal:03,barrett-degenaar-etal:15b,barrett-hilgen-etal:16}.    
%\pagebreak

This emphasizes the universality of dynamical systems theory and opens up  perspectives to better understand vision and its pathologies, and to help elaborating treatments.

\section*{Acknowledgements}
This paper is a with joint collaborations with I. Ampuero, R. Cofr\'e, L. Gil, D. Karvouniari, O. Marre, S. Picaud, S. Souhiel and I would like to thank them.
This work benefited from the support of the Ecole Doctorale des  Sciences et Technologies de l'Information et de la Communication de Nice-Sophia-Antipolis (EDSTIC), the Agence Nationale de la Recherche (ANR), project Trajectory \url{https://anr.fr/Project-ANR-15-CE37-0011}, and the Neuromod institute of the University C\^ote d'Azur.

%\bibliographystyle{ws-rv-van}
%\bibliography{odyssee,biblio_Dora}

%\blankpage
%\printindex[aindx]                 % to print author index
%\printindex                        % to print subject index

\end{document}